\documentclass[acmsmall,screen,nonacm]{acmart}
\usepackage{booktabs}
\usepackage{array}
\usepackage{graphicx}
\usepackage{listings}
\usepackage{mathpartir}
\usepackage[capitalise]{cleveref}
\graphicspath{{generated/figures/}}

\newcommand{\status}[1]{\textsf{\small[#1]}}
\newcommand{\den}[1]{[\![#1]\!]}
\newcommand{\lean}[1]{\texttt{#1}}
\newcommand{\nBaseBenchmarks}{50}

\newcommand{\nBaseBqcpFf}{17}

\newcommand{\nBaseQiskitOThreeMoreDyn}{6}

\newcommand{\nDeferRows}{188}
\newcommand{\nDeferCertified}{180}
\newcommand{\nDeferBnbMatch}{180}
\newcommand{\nDeferViolations}{0}
\newcommand{\nDeferChecked}{3046}
\newcommand{\nDeferUnchecked}{26}
\newcommand{\nDeferDeferWinsSixZeroZero}{9}
\newcommand{\nDeferCandRowsSixZeroZero}{21}

\newcommand{\nSeleneVariants}{220}
\newcommand{\nSeleneBenchmarks}{44}
\newcommand{\nSeleneShots}{44000}
\newcommand{\nSeleneExact}{39988}
\newcommand{\nSeleneVariantsExact}{199}

\newcommand{\nIbmSlope}{3.45}
\newcommand{\nIbmSlopeLo}{0.32}
\newcommand{\nIbmSlopeHi}{6.58}
\newcommand{\nIbmPairs}{27}

\newcommand{\nIbmKeptMean}{-2.95}
\newcommand{\nIbmKeptSe}{10.44}
\newcommand{\nIbmKeptLo}{-24.50}
\newcommand{\nIbmKeptHi}{18.60}

\newcommand{\nCsixbRounds}{15}

\newcommand{\nCsixbDropped}{4}
\newcommand{\nCsixbMean}{6.97}

\newcommand{\nCsixbLo}{-8.69}
\newcommand{\nCsixbHi}{22.63}
\newcommand{\nCsixbTOneNineP}{0.1156}
\newcommand{\nCsixbWilcoxonP}{0.8040}

\newcommand{\nCsixbModel}{19.22}
\newcommand{\nCsixbModelSource}{78.35}
\newcommand{\nCsixbModelDefer}{59.13}
\newcommand{\nCsixbPooledSource}{41.89}
\newcommand{\nCsixbPooledDefer}{36.69}
\newcommand{\nCsixbEffSource}{1.62}
\newcommand{\nCsixbEffDefer}{1.95}

\newcommand{\nSwitchBenchmarks}{65}

\newcommand{\nSwitchNoneTableTen}{4}
\newcommand{\nSwitchNoneTableMax}{16.0}
\newcommand{\nSwitchQubicTb}{14}

\newcommand{\nSwitchQickTb}{42}

\newcommand{\nSwitchQickTreeTen}{10}
\newcommand{\nSwitchQickTreeMax}{91.2}

\newcommand{\nQubicCond}{6.0}
\newcommand{\nQubicJump}{4.0}
\newcommand{\nQubicResid}{0.975}
\newcommand{\nQubicPoints}{312}
\newcommand{\nQubicRepSource}{151.1}
\newcommand{\nQubicRepChain}{21.3}
\newcommand{\nQubicRepSourceWorst}{155}
\newcommand{\nQubicRepChainWorst}{225}
\newcommand{\nQubicFiveSource}{150.9}
\newcommand{\nQubicFiveTree}{79.7}
\newcommand{\nFullPrograms}{42}
\newcommand{\nFullTabtreeMax}{0.3}
\newcommand{\nFullChainWins}{3}
\newcommand{\nFullSteaneMargin}{11.4}
\newcommand{\nFullSurfaceMarginLo}{5.9}
\newcommand{\nFullSurfaceMarginHi}{6.0}
\newcommand{\nFullRepFull}{91.2}
\newcommand{\nFullRepFullCases}{1024}
\newcommand{\nFullRepTwo}{83.5}
\newcommand{\nFullRepTwoCases}{56}
\newcommand{\nFullRepFour}{90.5}
\newcommand{\nFullRepFourCases}{386}
\newcommand{\nNoisePLo}{0.001}
\newcommand{\nNoisePHi}{0.1}

\newcommand{\nNoiseNoneTenLo}{4}

\newcommand{\nNoiseQickTenLo}{10}

\newcommand{\nNoiseRepBestLo}{90.9}
\newcommand{\nNoiseRepBestHi}{91.2}
\newcommand{\nNoiseRepTwoLoP}{91.2}
\newcommand{\nNoiseRepTwoHiP}{67.7}

\setcopyright{none}
\acmJournal{PACMPL}
\acmVolume{1}
\acmNumber{OOPSLA}
\acmYear{2027}

\title{Latency-Preserving Rewrites and Translation Validation for
Dynamic Quantum Circuits}

\author{Rylan Malarchick}
\email{rylan1012@gmail.com}
\affiliation{\institution{Embry-Riddle Aeronautical University}\city{Daytona Beach}\state{Florida}\country{USA}}

\begin{abstract}
A dynamic quantum circuit measures qubits during a run and chooses later
operations from the outcomes. These branches make the qubits wait:
the outcome must reach the classical controller, and the controller must
decide. We present qlin, a compiler that reduces this waiting time. qlin
moves operations out of branches without delaying the program, a
property proved in Lean~4. For error-correction decoders, which branch on
syndromes that are almost always zero, qlin compiles the likely
syndromes into one multiway branch and keeps the original program for
the rest. A cost model decides when a classically controlled gate should
become a quantum-controlled gate instead. On all benchmarks, qlin checks that the compiled program computes what its source computes,
and the checker is proved sound in Lean~4. At the published per-decision cost of the QICK
controller (\nSwitchQickTb~ns), \nSwitchQickTreeTen{} of
\nSwitchBenchmarks{} programs run at least 10\% faster than the best
baseline in our cost model, up to \nSwitchQickTreeMax\%, mostly because
one multiway branch replaces a long chain of \texttt{if} statements. On
the register-transfer logic of the QubiC controller, a compiled
distance-5 repetition decoder selects its correction in
\nQubicRepChain{} instead of \nQubicRepSource{} expected cycles. Timing
runs on an IBM device were too noisy to confirm or reject the model's
ranking.

\end{abstract}



\begin{document}
\maketitle

\section{Introduction}
\label{sec:intro}

A dynamic circuit measures a qubit, sends the outcome to a classical
controller, and uses it to choose the next operations. Error-correction
rounds, repeat-until-success loops, teleportation, and iterative phase
estimation all work this way. A classically controlled operation waits for its measurement to end and then for a feedforward latency
$t_\mathrm{ff}$. On IBM Heron devices $t_\mathrm{ff}$ is about
600~ns~\cite{shirgure2026dynamarq}, while a two-qubit gate on
ibm\_kingston takes 68~ns. The controller also spends time deciding a branch, and it can hold only a limited number of precompiled code
segments.

Compilers for dynamic circuits mostly remove branches. Branch-aware
quantum constant propagation (BQCP) removes branches whose condition is
known at compile time and proves its analysis sound on
paper~\cite{fulginiti2026bqcp}. Related passes simplify classically
controlled operations by symbolic simulation~\cite{fulginiti2026compile},
turn quantum operations into classical ones~\cite{remme2026dequant}, or
specialize code per branch~\cite{chen2025branch}. The branches that survive these passes are still decided at run time, in all shots.
Verified compilers also reach this setting. QPPM verifies in Lean a
compiler for Pauli-product-measurement programs with classical feedback
and certifies a depth bound for the candidates that it selects~\cite{ye2026qppm}, and the type system of HyQ guarantees that all circuits are generated before the program runs~\cite{sano2026hyq}.
Neither proves that a rewrite cannot delay a program whose timing
includes feedforward latency, and a rewrite that keeps the semantics can
make a program slower (\cref{sec:m0}).

This paper presents qlin, a compiler that reduces the time a dynamic
circuit spends on classical control. It has three parts. First, a normal
form moves operations out of branches (hoist and merge) without delaying
the program: for all moves in an ideal timing model, and for moves of single operations with a controller that decides one branch at a time, with
proofs in Lean. qlin checks the compiled output against its source,
and the checker is proved sound in Lean, so a bug in the Rust passes shows up
as a rejected output, not as a wrong program. Second, a fast path for likely outcomes.
Most of the probability of an error-correction round sits on low-weight
syndromes. qlin fixes a reference outcome $r$ and a budget $t$, and it
compiles a multiway branch (a \texttt{switch}) with one case per outcome within $t$ bits of $r$. Any other outcome takes an \emph{exit
edge} to the unchanged program. This is the hot and cold split of trace
scheduling~\cite{fisher1981trace} and deoptimization~\cite{holzle1992deopt},
with two differences: the profile comes from a noise model, and the exit edges sit before any specialized operation, so the fallback needs no
compensation code on a quantum state that the controller cannot read.
Third, a cost model chooses, per branch, between classical
feedforward and a quantum-controlled gate (deferred measurement).
Deferring can remove all runtime branches of teleportation and still be
slower on some devices, so the choice is a search over expected cost.

The fast path helps only when the controller spends time on a branch decision beyond the feedforward wait. We take that time from published
controllers, \nSwitchQubicTb~ns per control-flow instruction on
QubiC~2.0~\cite{xu2023qubic} and \nSwitchQickTb~ns per conditional jump on
QICK~\cite{stefanazzi2021qick}, and from our measurement of
\nIbmSlope~$\mu$s on ibm\_kingston. A controller can dispatch a multiway
branch by a jump table, by a decision tree on the bits, or by a chain of
compares, and we evaluate all three.

\paragraph{Contributions.}
\begin{enumerate}
\item A normal form whose hoist and merge steps never delay the
  program, at any position and in any sequence: for all moves in an ideal timing model, and for moves of single operations with a serial controller.
  The proofs are in Lean, and all side conditions are shown to be necessary
  (\cref{sec:m0}).
\item Translation validation of the normal-form and fast-path passes:
  all 976 outputs on our benchmarks pass, and the 66 normal-form outputs
  also pass a timing check in both models. The checkers are proved sound
  in Lean (\cref{sec:tv}).
\item A semantics over classical-quantum (cq) states with a
  density-matrix model, in which we prove the normal-form rules, the
  fast-path step, the Pauli sink (which moves Pauli corrections past
  Clifford gates), and the direct defer rule, with Lean's standard axioms
  only (\cref{sec:ir,sec:soundness}).
\item The fast-path construction, proved sound for any set of cases and
  therefore for all budgets and size limits (\cref{sec:fastpath}).
\item A branch-and-bound search over defer choices, which finds the
  exhaustive optimum on \nDeferBnbMatch{} of \nDeferCertified{} rows
  (\cref{sec:defer}).
\item An evaluation on \nBaseBenchmarks{} programs against BQCP, Qiskit,
  and TKET and on 16 error-correction programs of 13 to 22 qubits, with a
  cycle count on the logic of the QubiC controller, an accounting check
  against the Selene emulator, and timing runs on ibm\_kingston
  (\cref{sec:eval}).
\end{enumerate}
A label in brackets gives the support for a claim: \status{Lean-checked},
\status{exhaustive}, \status{empirical}, or \status{hand-derived}, and for
a number \status{modeled}, \status{emulated}, \status{RTL-simulated}, or
\status{measured}. \Cref{app:ledger} lists the claims with their evidence.

At the QICK cost with tree dispatch, \nSwitchQickTreeTen{} of
\nSwitchBenchmarks{} programs run at least 10\% faster than the best
baseline, up to \nSwitchQickTreeMax\% \status{modeled}. They are noisy
repetition, Steane, and surface codes, whose lookup decoders have many
branches on few bits. Most of the gain comes from replacing a chain of
\texttt{if}s by one multiway branch. Without the sink, a full switch over all outcomes comes within \nFullTabtreeMax\% of the fast path, and the
budget $t$ keeps most of the gain with a fraction of the cases. With no
controller time per branch, \nSwitchNoneTableTen{} programs gain 10\% or
more, at most \nSwitchNoneTableMax\%, all from the sink. On the
five-qubit code the best defer program is fastest. On ibm\_kingston,
three timing experiments could not separate the variants
(\cref{sec:eval}).

\section{Overview}
\label{sec:overview}

\Cref{fig:overview} (left) shows one round of the three-qubit bit-flip
repetition code. Qubits \texttt{q3} and \texttt{q4} collect the parities
of neighboring data qubits, and a lookup decoder applies one correction per nonzero syndrome. The decoder is three \texttt{if}s on the same
two bits. An \texttt{if} is a runtime branch: it waits for both
measurements, plus $t_\mathrm{ff}$, and a serial controller spends
$t_\mathrm{branch}$ deciding it.

\begin{figure}[t]
\centering
\begin{minipage}[t]{0.42\linewidth}
\begin{lstlisting}
qubits 5
bits 2
x q0
cx q0 q1
cx q1 q2
cx q0 q3
cx q1 q3
cx q1 q4
cx q2 q4
measure q3 -> c0
measure q4 -> c1
if c0 & !c1 { x q0 }
if c0 & c1 { x q1 }
if !c0 & c1 { x q2 }
\end{lstlisting}
\end{minipage}\hfill
\begin{minipage}[t]{0.54\linewidth}
\begin{lstlisting}
... (same prefix) ...
measure q3 -> c0
measure q4 -> c1
switch c0 c1 {
  case 0 0 { }
  case 0 1 { x q2 }
  case 1 0 { x q0 }
  default {
    if c0 & !c1 { x q0 }
    if c0 & c1 { x q1 }
    if !c0 & c1 { x q2 }
  }
}
\end{lstlisting}
\end{minipage}
\caption{Left: one round of the three-qubit repetition code with a
lookup decoder, in the qlin text format. Right: the fast path for the
reference outcome $r = 00$ and budget $t = 1$, as printed by
\texttt{qlin fastpath}. The default case is the exit edge. It runs the
source decoder unchanged.}
\label{fig:overview}
\Description{Two program listings side by side: the repetition-code round with three if statements, and its fast path with one switch statement.}
\end{figure}

\paragraph{Step 1: the normal form.} The compiler first puts the
program in a normal form: it folds constant conditions, hoists an op that
both arms of an \texttt{if} contain, and merges a common suffix. Here all three \texttt{if}s have an empty \texttt{else}, so nothing moves, and the
three corrections stay runtime decisions. \Cref{sec:m0} proves that these
moves never delay the program.

\paragraph{Step 2: classical or coherent control.} A correction
\texttt{if c0 \& !c1 \{x q0\}} could instead run as a coherent gate
controlled by \texttt{q3} and \texttt{q4}, with the measurements moved
after it. That gate is a three-qubit gate with a negated control. With
the assumed three-qubit gate time of our starting cost model, the search
of \cref{sec:defer} keeps all three corrections classical. On
\texttt{jeff/teleportation} the same search defers both corrections and cuts the
modeled expected latency from 2090 to 1136~ns.

\paragraph{Step 3: the fast path.} The three \texttt{if}s read only
\texttt{c0} and \texttt{c1}, and they write neither bit. They form
one \emph{dispatch group}. The fast path replaces the group by one
\texttt{switch} on \texttt{c0 c1} (\cref{fig:overview}, right). For a value within the budget, its case runs only the arms that the value
selects. Value \texttt{11} is outside $M_1$, so it takes the default,
which is the original group. The fallback starts from the state that the
source program would have at the same point, so no compensation code is
needed. A controller that spends $t_\mathrm{branch}=200$~ns per decision
now makes one decision instead of three if it has a jump table, and two
with a decision tree on the two bits. With the jump table the modeled
expected latency drops from 2372 to 1972~ns. With $t_\mathrm{branch}=0$ both programs take
1772~ns, because the switch waits for the same two measurements
\status{modeled}. This program has no
noise, so its syndrome is \texttt{00} in all shots. \Cref{sec:eval}
uses variants with injected errors.

\paragraph{Step 4: why it is sound.} The Lean theorem
\lean{fastpath\_step} (\cref{sec:fastpath}) says that the switch equals
``group, then rest'' when three conditions hold. The group conditions
read only the switch bits $G$. The group writes no bit of $G$. A case body equals the arms that its value selects, followed by the rest. All
three hold here. The theorem holds for any set of cases, so it holds for all budgets $t$ and all size limits. It is stated over any leaf
semantics that satisfies two axioms, and \cref{sec:ir} gives a
density-matrix model of gates, reset, and measurement in which both
axioms are theorems. The checker of \cref{sec:tv}, proved sound in
Lean, also validates the switches that the Rust code builds.

\section{Programs and Semantics}
\label{sec:ir}

\subsection{The Intermediate Representation (IR)}

A program has qubits $q_0, \dots, q_{n-1}$ and classical bits
$c_0, \dots, c_{m-1}$. A block is a list of operations (ops):
\begin{align*}
o \;::=\;& g\ \vec{q} \mid \texttt{measure}\ q \to b \mid \texttt{reset}\ q
  && \text{(leaves)}\\
  \mid\;& \texttt{if}\ e\ \{A\}\ \texttt{else}\ \{B\}
  \mid \texttt{loop max}\ K\ \{B\}\ \texttt{until}\ e
  \mid \texttt{switch}\ \vec{b}\ \{\texttt{case}\ \vec{v}_i\ \{B_i\}\ \dots\ \texttt{default}\ \{D\}\}
\end{align*}
Here $g$ is a gate with constant angles, $e$ is a Boolean expression over
bits, $A$, $B$, $B_i$, and $D$ are blocks, and $K$ bounds the iterations
of a loop that exits when $e$ holds after an iteration. A \texttt{switch}
runs the first case whose values equal the bits $\vec{b}$, else the
default. The IR has no static loops (the parser unrolls \texttt{repeat}),
and a bit-dependent angle is an \texttt{if}. \Cref{fig:overview} shows a
program in the text format.

\begin{definition}[Footprint]
The footprint $\mathit{fp}(o) = (Q, R, W)$ of an op is its set of qubits
$Q$, the bits $R$ that it reads, and the bits $W$ that it writes. A
measurement writes its bit. An \texttt{if}, a loop, and a
\texttt{switch} read the bits of their condition, and a compound op takes
the union over its parts. Two footprints are \emph{disjoint},
$(Q,R,W) \perp (Q',R',W')$, when $Q \cap Q' = W \cap W' = W \cap R' =
R \cap W' = \emptyset$.
\end{definition}

All moves in this paper use this one relation, and the Lean
definition copies the Rust one.

\subsection{Semantics over cq-states}

\begin{definition}[cq-state]
Let $D$ be an additive commutative monoid. A \emph{state} is a function
$s : \mathit{Assign} \to D$, where an assignment
$\beta \in \mathit{Assign}$ gives a value to all bits. For a condition
$c$, the restriction $s|_c$ keeps the branches where $c$ holds:
$s|_c(\beta) = s(\beta)$ if $c(\beta)$, else $0$.
\end{definition}

For a physical program, $D$ is the cone of unnormalized density
operators on $n$ qubits, and $s(\beta)$ is the quantum part on the branch
where the bits equal $\beta$. A block denotes the composition of its ops
from left to right, and
\begin{align*}
\den{\texttt{if}\ c\ A\ B}\,s &= \den{A}(s|_c) + \den{B}(s|_{\neg c}),\\
\den{\texttt{loop}\ K\ B\ c}\,s &= L_K(s), \quad
  L_0(s) = s,\quad L_{K+1}(s) = (\den{B}s)|_c + L_K\big((\den{B}s)|_{\neg c}\big),\\
\den{\texttt{switch}\ G\ ((v_1,B_1) :: cs)\ D}\,s &=
  \den{B_1}(s|_{G = v_1}) + \den{\texttt{switch}\ G\ cs\ D}(s|_{G \neq v_1}),
  \qquad \den{\texttt{switch}\ G\ []\ D} = \den{D}.
\end{align*}
A branch still in the loop after $K$ iterations stops there. This is the
truncated trace that the implementation enumerates.

\begin{definition}[Interpretation]
An interpretation $I$ gives a leaf an additive map on states and
satisfies two axioms:
\begin{description}
\item[A1 (locality)] A leaf commutes with $s \mapsto s|_c$ when $c$
  reads no bit that the leaf writes.
\item[A2 (disjointness)] Two leaves with disjoint footprints commute.
\end{description}
\end{definition}

Both axioms lift from leaves to all ops by induction on the op
(\lean{den\_restrict} and \lean{den\_comm}) \status{Lean-checked}. A
third fact follows from A1: an op keeps a bit at $0$ on the support of
the state when it does not write that bit (\lean{supp\_den}). Constant
folding uses this fact. qlin knows the value of a bit only
before any op writes it. It never infers a bit value from a quantum
state, which is the job of a propagation pass such as
BQCP~\cite{fulginiti2026bqcp}.

\subsection{A density-matrix model}
\label{sec:cq-model}

The results of \cref{sec:m0,sec:fastpath} hold for all interpretations.
This subsection gives the physical one and proves that it satisfies A1
and A2, so that those results apply to quantum channels without a
hand-checked step.

Fix $n$ qubits with computational basis $\{0,1\}^n$, and let $D$ be the
complex $2^n \times 2^n$ matrices. For a set of qubits $Q$, a matrix $k$
is \emph{local to $Q$} when $k_{xy}$ depends only on the values of $x$
and $y$ on $Q$. Its \emph{lift} acts as $k$ on $Q$ and as the identity
elsewhere:
\[
(\mathrm{lift}_Q\, k)_{xy} = \begin{cases} k_{xy} & \text{if } x \text{ and } y \text{ agree outside } Q,\\ 0 & \text{otherwise.}\end{cases}
\]
A Kraus map on $Q$ with operators $K_1, \dots, K_r$ is
$\rho \mapsto \sum_i (\mathrm{lift}_Q K_i)\,\rho\,(\mathrm{lift}_Q K_i)^\dagger$.
The leaves are:
\begin{itemize}
\item a gate with matrix $U$ local to its qubits $Q$: the Kraus map with
  one operator $U$, applied on all branches;
\item \texttt{reset q}: the Kraus map on $\{q\}$ with operators
  $|0\rangle\langle 0|$ and $|0\rangle\langle 1|$, on all branches;
\item \texttt{measure q -> b}: with $P_v$ the projector on
  $q = v$,
  \[
  \den{\texttt{measure}\ q \to b}\,s\,(\beta) = P_{\beta(b)}\big(s(\beta[b \mapsto 0]) + s(\beta[b \mapsto 1])\big)P_{\beta(b)}.
  \]
\end{itemize}
The measurement collects both old values of $b$ and keeps the part of
the state where qubit $q$ agrees with the new value of $b$.

\begin{theorem}[A1 and A2 for density matrices]
\label{thm:cq}
In this model all leaves satisfy A1, and any two leaves with disjoint
footprints commute (A2). (\lean{Spec.sem\_restrict},
\lean{Spec.sem\_comm}, and the interpretation \lean{CQ.interp})
\status{Lean-checked}
\end{theorem}

The core of the proof is that lifts of matrices local to disjoint qubit
sets commute (\lean{lift\_comm}). Entry $(x,z)$ of
$\mathrm{lift}_P\,a \cdot \mathrm{lift}_Q\,b$ has one nonzero summand,
at the basis state that agrees with $z$ on $P$ and with $x$ elsewhere.
Both products equal $a_{xz} b_{xz}$ when $x$ and $z$ agree outside
$P \cup Q$, and $0$ otherwise. Two measurements of different qubits into different bits
commute because the two updates of $\beta$ commute and the projectors
commute. As a check that the model is faithful, the Kraus operators of
measurement and of reset satisfy $\sum_i K_i^\dagger K_i = 1$
(\lean{meas\_complete}, \lean{reset\_complete}), so both are trace
preserving \status{Lean-checked}.

The Lean interpretation takes a table that assigns one of these operations to a leaf id. A leaf whose declared footprint does not cover its
operation denotes the identity, which keeps both axioms true for all declared footprints. A leaf whose footprint is its own, as in all programs that the Rust code builds, denotes its operation
(\lean{interp\_leaf}).

\section{The Normal Form and a Latency Theorem}
\label{sec:m0}

\subsection{Replay latency}

qlin scores a program by its expected replay latency under a cost model
$C$, which gives a duration to a leaf and the feedforward latency
$t_\mathrm{ff}$. An \emph{outcome record} $R$ gives the outcome of the
$k$-th measurement of a bit. The replay runs the trace that $R$
selects and keeps a state $\sigma$ with five parts: the time
$\mathit{ready}(q)$ at which qubit $q$ is free, the end time
$\mathit{mEnd}(b)$ of the latest measurement of bit $b$, the current bit
values, a cursor per bit into $R$, and a stop flag. An op runs with the set $\gamma$ of \emph{gate bits}: the condition bits of all enclosing
\texttt{if}, \texttt{switch}, and loop exit. A leaf with footprint
$(Q, R, W)$ and duration $d$ finishes at
\[
\mathit{fin} = \max\Big(\max_{b \in \gamma}\big(\mathit{mEnd}(b) + t_\mathrm{ff}\big),\ \max_{q \in Q} \mathit{ready}(q)\Big) + d,
\]
which becomes the ready time of all $q \in Q$ and the measurement end of all $b \in W$. A measurement reads its outcome from $R$ and advances the
cursor of its bit. An \texttt{if} runs its taken arm with
$\gamma \cup \mathit{bits}(c)$. A loop that does not exit within $K$
iterations sets the stop flag, and the replay skips all later ops. The
makespan is the largest ready time. The expected latency weights the
makespan of a record by its probability, which comes from one
state-vector simulation of the source (at most 12 qubits).

We call this the \emph{ideal} model. The \emph{block} model adds a
serial controller. An executed branch takes $t_\mathrm{branch}$ of controller time and stalls the qubits in its footprint until the decision, taken or not. Numbers from these models are labeled
\emph{modeled}.

\subsection{The normal form}

The normal form (M0 in our tables) applies four moves:
\begin{itemize}
\item \emph{fold} a condition that is constant on the known bits, and
  keep the live arm;
\item \emph{hoist} an op $x$ that both arms contain in front of the
  \texttt{if};
\item \emph{merge} an op $x$ that both arms contain to after the
  \texttt{if};
\item \emph{unroll} the first iteration of a loop.
\end{itemize}
Hoist and merge carry side conditions. Write $p_A$ and $p_B$ for the ops
that $x$ moves past in the two arms, $c$ for the condition, and $\gamma$ for
the gate bits of the \texttt{if}:
\begin{mathpar}
\inferrule[Hoist]
  {\forall y \in p_A \cup p_B.\ \mathit{fp}(x) \perp \mathit{fp}(y) \\
   W(x) \cap (\mathit{bits}(c) \cup \gamma) = \emptyset}
  {\texttt{if}\ c\ (p_A \mathbin{+\!\!+} x :: A)\ (p_B \mathbin{+\!\!+} x :: B)
   \;\leadsto\; x;\ \texttt{if}\ c\ (p_A \mathbin{+\!\!+} A)\ (p_B \mathbin{+\!\!+} B)}

\inferrule[Merge]
  {\forall y \in s_A \cup s_B.\ \mathit{fp}(x) \perp \mathit{fp}(y) \\
   \forall y \in s_A \cup s_B.\ W(y) \cap \gamma = \emptyset}
  {\texttt{if}\ c\ (A \mathbin{+\!\!+} x :: s_A)\ (B \mathbin{+\!\!+} x :: s_B)
   \;\leadsto\; \texttt{if}\ c\ (A \mathbin{+\!\!+} s_A)\ (B \mathbin{+\!\!+} s_B);\ x}
\end{mathpar}
The conditions on $\gamma$ matter only for latency. The semantic rules
need only the disjointness conditions and, for hoist,
$W(x) \cap \mathit{bits}(c) = \emptyset$ (\cref{sec:soundness}).

The \emph{dynamic core} is the set of leaves that stay inside a runtime
branch after all moves apply. It holds the unmatched ops of the dynamic \texttt{if}s and the leaves of a loop body that repeats on a
measured bit. The core leaves and the number of dynamic conditions are
the program metrics of \cref{tab:baselines}.

\subsection{The latency theorem}

\begin{theorem}[C1: hoist and merge never delay the replay]
\label{thm:c1}
For all cost models, records $R$, gate bits $\gamma$, and start state
$\sigma$ that is not stopped: if one hoist or merge step under the side
conditions above turns op $o$ into block $o'$, and the replay of $o$ does
not stop, then the replay state after $o'$ is no later than the replay
state after $o$. That is, it has the same bit values, cursors, and stop
flag, and no ready time or measurement end is later than the original.
(\lean{hoist\_latency}, \lean{merge\_latency}) \status{Lean-checked}
\end{theorem}

The proof rests on two lemmas. The replay is monotone in its start state
and in its gate bits (\lean{run\_mono}). An exchange lemma also holds: a
block that writes no gate bit of a disjoint block can move in
front of it without delay. A pointwise smaller state has a smaller
makespan (\lean{makespan\_le}). The expected latency is a sum over
records with nonnegative weights, so it does not grow either
(\lean{hoist\_expected\_latency}, \lean{merge\_expected\_latency})
\status{Lean-checked}. The theorem covers the ideal model on
records with no truncated loop. The expected forms score both programs
on the outcome records of the source.

\begin{theorem}[Closure of C1]
\label{thm:c1closure}
Let $P \Rightarrow_E Q$ hold when $Q$ follows from $P$ by hoist and merge
steps at any position: in a block, in an arm of an \texttt{if}, in a loop
body, or in a case or the default of a \texttt{switch}, and by any
sequence of such steps. Here $E$ bounds the enclosing gate bits (an arm
adds its condition bits, a case adds the switch bits), and all steps meet the side conditions above with $E$ in the place of $\gamma$.
If $P \Rightarrow_E Q$, then for all gate bits $\gamma \subseteq E$ and
all start states $\sigma' \le \sigma$, if the replay of $P$ from
$\sigma$ does not stop, the replay state after $Q$ from $\sigma'$ is no
later than the replay state after $P$ from $\sigma$.
(\lean{m0Steps\_latency}, \lean{m0Steps\_expected\_latency})
\status{Lean-checked}
\end{theorem}

The proof is an induction on the derivation, with \cref{thm:c1} at the
root and the monotonicity of the replay in all contexts. A truncated loop stops the replay, and
the latency report gives the probability of such records apart.

\paragraph{With a serial controller.} The block model of
\cref{sec:eval} adds a controller that decides one branch at a time.
A hoisted \texttt{if} can then add latency
(\lean{hoist\_block\_can\_delay}): on one program the normal form
raises the expected latency from 2078 to 2208~ns \status{modeled}.

\begin{theorem}[C1 in the block model]
\label{thm:c1block}
If all hoisted or merged ops are leaves, and a merged leaf writes no
enclosing gate bit and no condition bit, then \cref{thm:c1} and
\cref{thm:c1closure} hold in the block model.
(\lean{hoist\_block\_leaf}, \lean{merge\_block\_leaf},
\lean{m0StepsK\_latency}) \status{Lean-checked}
\end{theorem}

\paragraph{Why the conditions on $\gamma$.} Without them a hoist can
delay a program. In \cref{fig:c1} the outer \texttt{if} reads
\texttt{c0}, and both inner arms end with \texttt{measure q0 -> c0}.
Hoisting that measurement in front of the inner \texttt{if} keeps the
semantics, but \texttt{x q1} now waits for the new measurement of
\texttt{c0}, because an op waits for the latest measurement of all enclosing condition bits. The expected latency rises from 1532 to
1690~ns \status{modeled}. A merge has the mirror problem: an op that
moves past a new write of an enclosing bit waits for that write (1348 to
1506~ns on a second program). This rule of the replay is conservative: a
controller that latched the condition at the decision would not wait.

\begin{figure}[t]
\centering
\begin{minipage}{0.47\linewidth}
\begin{lstlisting}
h q0
measure q0 -> c0
h q2
measure q2 -> c1
if c0 {
  if c1 { x q1  measure q0 -> c0 }
  else  { measure q0 -> c0 }
}
\end{lstlisting}
\end{minipage}\hfill
\begin{minipage}{0.47\linewidth}
\begin{lstlisting}
h q0
measure q0 -> c0
h q2
measure q2 -> c1
if c0 {
  measure q0 -> c0
  if c1 { x q1 }
}
\end{lstlisting}
\end{minipage}
\caption{A hoist that delays a program (3 qubits, 2 bits). Left:
source, 1532~ns expected. Right: after a hoist that the condition on
$\gamma$ forbids, 1690~ns (modeled). The normal form instead keeps the
measurement in the inner arms and merges it after the inner
\texttt{if}.}
\label{fig:c1}
\Description{Two program listings: the nested-branch source and the same program with the measurement hoisted.}
\end{figure}

Negative theorems show that the conditions on $\gamma$ are necessary: without one, instances of the two programs above refute both steps
(\lean{hoist\_latency\_needs\_henc}, \lean{merge\_latency\_needs\_henc})
\status{Lean-checked}. No measurement inside a branch of the
\nBaseBenchmarks{} benchmark files writes a bit that an enclosing branch
reads. An exhaustive check compares the normal form with all prefixes of the hoist and merge sequences at the top-level \texttt{if}s, on all benchmarks with at most 12 qubits \status{exhaustive}.

\section{Expected Cost and the Defer Choice}
\label{sec:defer}

\subsection{The objective}

The compiler scores a program by its expected replay latency over the outcome records. Two choices set it: which \texttt{if}s to defer,
which the search of this section makes, and the fast-path budget $t$,
which the evaluation sweeps under a limit $S$ on the number of case
blocks. For a program with one dispatch group the expected cost splits
as
\[
\mathbb{E}[\mathrm{cost}] = \sum_{\beta \in M_t} p(\beta)\, C(F, \beta)
 + \sum_{\beta \notin M_t} p(\beta)\,\big(C(\mathrm{exit}) + C(G, \beta)\big),
\]
where $F$ is the fast path, $G$ the fallback, and $p$ the outcome
distribution. With several groups an outcome can leave the budget at a
later group, and the replay accounts for it. The objective is cost, not
the size of the dynamic core.
The deferred-measurement rewrite (defer) replaces a classically
controlled gate by a quantum controlled gate and moves the measurement
later. It removes a feedforward wait, and it costs a multi-qubit gate and
sometimes a fresh qubit. With defer, teleportation keeps no runtime
branch \status{hand-derived}, but the coherent program is not faster on all devices, so the cost model decides.

\subsection{The defer rule}

A candidate is a top-level dynamic \texttt{if}. Its condition bits were last written by a top-level measurement, and its arms hold only
gates with a controlled form. A deferred \texttt{if} reads its bits from
a control qubit in one of two forms. The \emph{direct} form uses the
measured qubit itself, and the measurement moves to after its last user.
It needs no other op on that qubit, no classical reader of the bit, and
no new write of the bit in between. The \emph{copy} form uses a fresh
ancilla: \texttt{cx q anc} goes just before the measurement, and the
gates use the ancilla.

\begin{theorem}[Direct defer, one control bit]
\label{thm:defer}
In the density-matrix model, let $U$ be any matrix local to a qubit set
$T$ with $q \notin T$, and let $C_v U$ be $U$ on $T$ controlled by qubit
$q$ on value $v$ (the identity on $T$ when qubit $q$ holds $\neg v$). Then
\[
\den{\texttt{measure}\ q \to b;\ \texttt{if}\ (b = v)\ \{U\}}
= \den{C_v U;\ \texttt{measure}\ q \to b}.
\]
(\lean{defer\_cq}) \status{Lean-checked}
\end{theorem}

The proof is an identity of matrices. With $P_w$ the projector on
$q = w$, the product $P_w \cdot C_v U$ equals $U \cdot P_v$ when $w = v$
and $P_w$ otherwise (\lean{proj\_ctrl}). Moving the measurement further,
past ops that touch neither $q$ nor $b$, is A2. The theorem covers the
direct form with one control bit and one gate in the arm. Four cases stay outside
Lean: the copy form (which needs a partial trace over the ancilla),
conditions on several bits, XOR conditions, and arms with several gates.
Simulation checks them \status{empirical}, and symbolic evaluation checks two programs \status{Wolfram-checked} (\cref{sec:soundness}).

\subsection{Search}

The search keeps a candidate classical or defers it. The objective is expected
replay latency, with ties broken on fewer added multi-qubit gates, then
fewer added qubits. For a partial assignment, the lower bound keeps the
undecided \texttt{if}s classical but \emph{relaxed}: they skip the
feedforward wait. On all records, relaxed is no slower than classical:
a relaxed \texttt{if} runs its arms under fewer gate bits, and the
replay is monotone in its gate bits (\lean{run\_mono}; Lean does not
model relaxed \texttt{if}s). It is also no slower than deferred:
the base gate is no slower than its controlled form, it runs only when
its condition holds, and the measurement does not move. This second
part is hand-derived, and an exhaustive search checks the bound on all prefixes \status{exhaustive}. Branch and bound prunes with it.

\paragraph{Results.} Over \nDeferRows{} benchmark and $t_\mathrm{ff}$
rows (0, 150, 600, and 2400~ns, HERON\_LIKE values), exhaustive search
certifies \nDeferCertified{} rows. Branch and bound finds the exhaustive
optimum on \nDeferBnbMatch{} of them, with \nDeferViolations{} bound
violations \status{exhaustive}. At $t_\mathrm{ff}=600$~ns the simulation
check passed on \nDeferChecked{} scored search leaves. It did not run on
\nDeferUnchecked{} leaves, which need more than 12 qubits with their copy
qubits. \Cref{tab:defer} lists the rows with candidates. Defer beats the
normal form on \nDeferDeferWinsSixZeroZero{} of the
\nDeferCandRowsSixZeroZero{} benchmarks with candidates at this
$t_\mathrm{ff}$. On \texttt{jeff/teleportation} the best program defers both
corrections. It takes 1136~ns against 2090~ns for the source, with two
added two-qubit gates and no new qubit \status{modeled}. The floor column
is the relaxed bound at the root.

\begin{table}[t]
\centering
\scriptsize
\begin{tabular}{lrrrrrrrr}
\toprule
benchmark & cand. & source & M0 & best & floor & deferred & +2q & +q \\
\midrule
\texttt{dynamarq/cnot\_ladder4} & 3 & 2317 & 2317 & 1236 & 1080 & 3 & 6 & 0 \\
\texttt{dynamarq/fanout4} & 3 & 2399 & 2399 & 1376 & 1220 & 3 & 8 & 0 \\
\texttt{dynamarq/five\_qubit\_code} & 5 & 4278 & 4278 & 3565 & 3213 & 5 & 5 & 0 \\
\texttt{dynamarq/five\_qubit\_code\_noisy} & 5 & 4189 & 4189 & 3565 & 3211 & 5 & 5 & 0 \\
\texttt{dynamarq/ghz5} & 4 & 2310 & 2310 & 1240 & 998 & 4 & 10 & 0 \\
\texttt{dynamarq/ipe3} & 3 & 5860 & 5860 & 5860 & 5860 & 0 & 0 & 0 \\
\texttt{dynamarq/long\_range\_cnot3} & 2 & 2074 & 2074 & 1136 & 1016 & 2 & 3 & 0 \\
\texttt{dynamarq/repetition3\_0} & 3 & 1172 & 1172 & 1172 & 1172 & 0 & 0 & 0 \\
\texttt{dynamarq/repetition3\_0\_noisy} & 3 & 1376 & 1376 & 1376 & 1174 & 0 & 0 & 0 \\
\texttt{dynamarq/repetition3\_1} & 3 & 1340 & 1340 & 1340 & 1340 & 0 & 0 & 0 \\
\texttt{dynamarq/repetition3\_1\_noisy} & 3 & 1541 & 1541 & 1541 & 1342 & 0 & 0 & 0 \\
\texttt{hand/mbqc\_chain4} & 4 & 4632 & 4632 & 1604 & 1108 & 4 & 7 & 0 \\
\texttt{hand/repetition3} & 3 & 1172 & 1172 & 1172 & 1172 & 0 & 0 & 0 \\
\texttt{hand/teleportation} & 2 & 1466 & 1466 & 1104 & 1000 & 2 & 2 & 0 \\
\texttt{jeff/iqft\_3} & 3 & 5544 & 5544 & 5544 & 5544 & 0 & 0 & 0 \\
\texttt{jeff/iqft\_5} & 10 & 9320 & 9320 & 9320 & 9320 & 0 & 0 & 0 \\
\texttt{jeff/iqft\_7} & 21 & 13160 & 13160 & 13160 & 13160 & 0 & 0 & 0 \\
\texttt{jeff/iqpe\_3} & 3 & 5844 & 5844 & 5844 & 5844 & 0 & 0 & 0 \\
\texttt{jeff/iqpe\_5} & 10 & 9820 & 9820 & 9820 & 9820 & 0 & 0 & 0 \\
\texttt{jeff/iqpe\_7} & 21 & 13684 & 13684 & 13684 & 13684 & 0 & 0 & 0 \\
\texttt{jeff/teleportation} & 2 & 2090 & 2090 & 1136 & 1000 & 2 & 2 & 0 \\
\bottomrule
\end{tabular}

\caption{Defer search at $t_\mathrm{ff}=600$~ns, HERON\_LIKE values
(modeled, expected latency in ns). ``cand.'' is the number of
candidates, ``floor'' the root lower bound, ``+2q'' and ``+q'' the added
two-qubit gates and qubits of the best program.}
\label{tab:defer}
\end{table}

\section{The Fast Path: Speculation over Outcomes}
\label{sec:fastpath}

\subsection{Budget and construction}

Fix a reference outcome $r$: for a bit, the more likely value at that
point of the program (ties go to 0). The \emph{weight} of an assignment
is the number of its bits that differ from $r$. The budget $M_t$ is the
set of outcomes of weight at most $t$, and for $m$ bits
$|M_t| = \sum_{k \le t} \binom{m}{k}$ \status{Wolfram-checked for
$m \le 12$}.

A \emph{dispatch group} is a maximal run of consecutive top-level
dynamic \texttt{if}s whose condition bits $G$ are all written before the
run and not written inside it. For a group $g$, the rest $\rho$ of the
program, and a remaining budget $k$, the fast path is
\[
F_k(g; \rho) = \texttt{switch}\ G\ \big\{\ \texttt{case}\ v :
  \mathit{sel}_v(g) \mathbin{+\!\!+} F_{k - |v|}(\rho)
  \ \big|\ v \in \{0,1\}^G,\ |v| \le k\ \big\}\ \texttt{default}\ (g \mathbin{+\!\!+} \rho),
\]
where $|v|$ is the weight of $v$ and $\mathit{sel}_v(g)$ replaces the \texttt{if}s of $g$ by the arm that $v$ selects. $F_k$ of a program with
no further group is the program itself. A group stays generic when it
has too many bits, or when the tree would pass the size limit $S$ of
precompiled case blocks (4096 in our runs).

The default case is the exit edge, and it is the deoptimization point.
In a classical JIT compiler, deoptimization must rebuild the frames of
the unoptimized code from the state of the optimized
code~\cite{holzle1992deopt}. Here the default runs before any specialized
op of its group, so the state at the exit edge is the state that the
source program has at the same point. The fallback is the unchanged
source, and the compensation code is empty. The price is that the fast
path cannot speculate across a group boundary: a repair on exit would
have to act on a quantum state that the controller cannot read.

In the block model, the source takes one controller decision per
\texttt{if}. A \texttt{switch} takes one decision if the controller has a
jump table, one per switch bit if it dispatches with a decision tree on
the bits, and one per compare, up to the matching case, if it tests the
cases in order. The fast path saves time when these decisions are fewer
than the \texttt{if}s that it replaces. A lookup decoder for $m$
syndrome bits has up to $2^m$ \texttt{if}s, and a decision tree needs
$m$ decisions, so the saving survives without a jump table. In the ideal model a branch costs only
its feedforward wait, and a switch waits for the same bits, so the
dispatch by itself saves no time \status{hand-derived}. It can cost a
little: the code under the switch waits for all group bits, and at
$t_\mathrm{branch}=0$ the fast path is slower than the normal form on 8
programs, by at most 0.99\% \status{modeled}.

\subsection{Soundness of the dispatch step}

\begin{theorem}[Dispatch step]
\label{thm:fastpath}
For all interpretations, let $g$ be a block whose \texttt{if}
conditions read only bits in $G$ and which writes no bit of $G$. Let
$\mathit{cs}$ be any list of cases such that for all cases $(v, B)$ the body satisfies $\den{B} = \den{\mathit{sel}_v(g) \mathbin{+\!\!+}
\rho}$. Then
\[
\den{\texttt{switch}\ G\ \mathit{cs}\ (g \mathbin{+\!\!+} \rho)} = \den{g \mathbin{+\!\!+} \rho}.
\]
(\lean{fastpath\_step}) \status{Lean-checked}
\end{theorem}

Case $v$ runs only on branches where the bits of $G$ equal $v$. On those
branches all \texttt{if}s of the group take the arm that $v$ selects,
because the group writes no bit of $G$ (A1, lifted). The cases and the
default split the state, and the denotation is additive, so the sum
equals the group. The theorem quantifies over the list of cases, so it
covers all budgets, references, and size limits. The case
body of $F_k$ is again a fast path. A relation \lean{FastPathOf}
describes the programs that the construction can build, and induction
over it gives $\den{F_t(P)} = \den{P}$ for the whole tree
(\lean{fastpath\_sound}) \status{Lean-checked}. A negative theorem refutes the step when the group
writes a dispatch bit (\lean{fastpath\_needs\_hw}) \status{Lean-checked}.

\subsection{Pauli sink}

A correction \texttt{if} whose arms are products of $X$, $Y$, and $Z$ can
move forward. It passes an op with a disjoint footprint unchanged. It
passes a Clifford gate on its qubits, and its arms are conjugated through the gate: $C P = \varphi P' C$ with $\varphi = \pm 1$. The table covers
\texttt{h}, \texttt{s}, \texttt{sdg}, \texttt{sx}, \texttt{x}, \texttt{y},
\texttt{z}, \texttt{cx}, \texttt{cy}, \texttt{cz}, and \texttt{swap}, and its entries are Lean theorems over $2 \times 2$ or $4 \times 4$ complex
matrices. After the sink the Clifford ops no longer wait for the
decision of the correction, which is the source of the small gains at
$t_\mathrm{branch}=0$ in \cref{sec:eval}.

\begin{theorem}[Sink step in the density-matrix model]
\label{thm:sink}
Let $C$, $P$, and $P'$ be one-qubit matrices with $CP = \varphi P'C$ for
some $\varphi = \pm 1$, placed on qubit $q$. Then
$\den{P;\ C} = \den{C;\ P'}$. The same holds for a two-qubit $C$ on
qubits $a \neq b$ and a Pauli leaf on both of them, with
$C (A_1 \otimes A_2) = \varphi (B_1 \otimes B_2) C$.
(\lean{sink1\_cq}, \lean{sink2\_cq}, instances \lean{sink\_h\_cq} and
\lean{sink\_cx\_cq}) \status{Lean-checked}
\end{theorem}

The sign drops out of the channel because
$(\varphi K)\rho(\varphi K)^\dagger = |\varphi|^2 K \rho K^\dagger$
(\lean{chan\_sign}). Together with the rule \lean{sink\_past\_clifford},
which moves a whole Pauli \texttt{if} past $C$ given the arm facts, this
closes the sink step for quantum channels. A table entry holds for all nonzero multiples of $C$ (\lean{conj\_smul}), so the table can use
$\sqrt{2}\,H$ and $2\,\mathit{SX}$ and avoid square roots.

\paragraph{Size.} Without the sink, the quantum error correction (QEC)
benchmarks fit in $S=4096$ for all $t \le 6$. The sink splits a group
into several, and its variants of Steane, of the repetition codes of
distance 7 to 11, and of the surface codes reach the limit at $t \ge 3$.
\texttt{iqft\_7} and \texttt{iqpe\_7} reach it at $t \ge 4$. Groups past
the limit stay generic \status{empirical}.

\section{Mechanization and Checking}
\label{sec:soundness}

We check soundness at three levels. Lean~4~\cite{moura2021lean4} with
Mathlib~\cite{mathlib2020} proves that the rewrite \emph{rules} preserve
the denotation of \cref{sec:ir}. The checker of \cref{sec:tv}, also
proved sound in Lean, validates the programs that the Rust code outputs against their inputs. State-vector and stabilizer simulation compares outputs with sources on all benchmarks up to 22 qubits but one, a non-Clifford program with 14
qubits.

\subsection{What Lean proves}

The development has 4846 lines, and 6585 with the checkers of
\cref{sec:tv}. A
list of 73 theorems
(\cref{tab:lean} shows the main ones) is checked by a script: they depend only on \texttt{propext}, \texttt{Classical.choice}, and
\texttt{Quot.sound}, and no theorem uses \texttt{sorry}. The rule theorems quantify over all interpretations $I$ and all states $s$. The
\lean{\_cq} theorems instantiate them, or prove new facts, in the
density-matrix model of \cref{sec:cq-model}.

\begin{table}[t]
\centering
\small
\begin{tabular}{p{0.3\linewidth}p{0.62\linewidth}}
\toprule
theorem & statement \\
\midrule
\lean{den\_restrict}, \lean{den\_comm} & A1 and A2 lift from leaves to all ops. \\
\lean{fold\_true}, \lean{fold\_false} & A condition that is constant whenever the known bits are $0$ selects its arm, on states whose support has those bits at $0$. \\
\lean{hoist\_sound}, \lean{merge\_sound} & The semantic hoist and merge rules (\cref{sec:m0}). \\
\lean{unroll\_sound} & A loop with fuel $K+1$ equals its body, then an \texttt{if} on the exit that runs the loop with fuel $K$. \\
\lean{fastpath\_step} & \Cref{thm:fastpath}, for any set of cases. \\
\lean{sink\_past\_clifford} & A Pauli \texttt{if} followed by $C$ equals $C$ followed by the conjugated \texttt{if}, given the fact for both arms. \\
\lean{conj\_h} \dots\ \lean{conj\_swap} & Table entries: $C P = \varphi P' C$ with $\varphi = \pm 1$. \\
\lean{hoist\_latency}, \lean{merge\_latency} & \Cref{thm:c1} (C1). \\
\lean{m0Steps\_latency} & \Cref{thm:c1closure}: C1 at any position and for sequences of steps. \\
\lean{lift\_comm}, \lean{chan\_comm} & Lifts, and Kraus maps, on disjoint qubits commute. \\
\lean{Spec.sem\_restrict}, \lean{Spec.sem\_comm} & \Cref{thm:cq}: A1 and A2 for gates, reset, and measurement. \\
\lean{meas\_complete}, \lean{reset\_complete} & $\sum_i K_i^\dagger K_i = 1$ for measurement and reset. \\
\lean{sink1\_cq}, \lean{sink2\_cq} & \Cref{thm:sink}: the sink step for channels. \\
\lean{defer\_cq} & \Cref{thm:defer}: direct defer with one control bit. \\
\lean{hoist\_sound\_cq}, \lean{merge\_sound\_cq}, \lean{fastpath\_step\_cq} & The rule theorems in the density-matrix model. \\
\lean{checkBlock\_sound}, \lean{checkFP\_sound}, \lean{checkLat\_sound} & \Cref{thm:tv}: the translation-validation checkers. \\
\lean{m0StepsK\_latency}, \lean{checkLatLeaf\_soundK} & \Cref{thm:c1block} and its checker; \lean{hoist\_block\_can\_delay}: a hoisted \texttt{if} can add latency. \\
\bottomrule
\end{tabular}
\caption{The main Lean theorems. The checked list has 73 entries,
including the negative theorems.}
\label{tab:lean}
\end{table}

\paragraph{Negative theorems.} For all side conditions, a Lean theorem refutes the rule without it, by a concrete counterexample. The
counterexamples use an interpretation over $D = \mathbb{N}$ in which one
leaf sets bit 0 and another doubles the weight. The refuted rules are
hoist without the condition-bit rule, hoist or merge past an op that is
not disjoint, a dispatch whose group writes a dispatch bit, and the two
latency steps without the enclosing-bit rules.

\paragraph{Trusted base.} A reader must trust four things beyond the
Lean kernel, the checker's JSON parser, the Lean compiler that runs the checkers on the jobs, and the Rust exporter \texttt{src/tv.rs}, which must copy the footprints of all leaves. The first is that the Lean IR copies the control structure
of the Rust IR. The second is that the density-matrix model states the
textbook semantics of gates, reset, and Z-basis measurement on a cq
state. Its definitions are in \texttt{CQ.lean}, and the completeness
theorems above check one property of them. The third is that the timing
semantics copies the replay code (\texttt{Latency.lean} names the Rust
functions that it copies), except that the replay skips the injected
noise operations of the \texttt{\_noisy} programs, which Lean treats as
ordinary operations. The fourth is the gap between rules and code. For the normal-form and
fast-path passes the checker of \cref{sec:tv} closes it for the programs we compiled. For the sink and defer passes the simulation checks cover it.

\paragraph{How the proofs were made.} For the rule and latency theorems
we wrote all statements by hand, typechecked it with \texttt{sorry}, and
committed it before any proof. The Aristotle
prover~\cite{achim2025aristotle} filled those proofs in four rounds, and
we graded the rounds locally. A script checks that these statements are unchanged from their committed drafts, so no statement was
weakened to make a proof pass. The density-matrix files, the closure and
block-model theorems, and the checkers were written later, statement and
proof together, without Aristotle. CI repeats the build and the axiom
check.

\subsection{Translation validation}
\label{sec:tv}

The rule theorems do not show that the Rust code applies only those
rules. Translation validation closes this gap one program at a time:
instead of verifying the compiler, a checker verifies its outputs~\cite{pnueli1998tv,necula2000tv}. \texttt{Check.lean} defines a
syntactic copy of the IR with bit expressions and injectively coded
leaves, and two Boolean checkers. \lean{checkBlock} decides whether a
target block follows from a source block by fold steps on known-zero
bits, hoists, merges, and congruence into arms. It decides all semantic side conditions on the syntax. A second checker, \lean{checkLat}, accepts the
same derivations with one more argument, a bound $E$ of the enclosing
gate bits, and it also decides the latency side conditions of
\cref{thm:c1}: a hoisted op writes no bit of $E$, and the ops that a
merged op moves past write no bit of $E$. \lean{checkFP} decides whether a target
is a fast path of its input: equal programs, an equal head, or one
dispatch whose default is the input and whose case bodies are, again,
fast paths of the arms their values select. All three are proved sound:

\begin{theorem}[Checker soundness]
\label{thm:tv}
If $\lean{checkBlock}\ f\ W\ P\ Q = \mathit{true}$, then $\den{P}\,s =
\den{Q}\,s$ for all interpretations and all states $s$ whose support
has the bits of $W$ at $0$. If $\lean{checkFP}\ f\ P\ T = \mathit{true}$,
then $\den{T} = \den{P}$ for all interpretations.
If $\lean{checkLat}\ f\ W\ E\ P\ Q = \mathit{true}$, then for all gate
bits $\gamma \subseteq E$ and all start states $\sigma' \le \sigma$ with
the bits of $W$ at $0$, if the replay of $P$ from $\sigma$ does not stop,
the replay state after $Q$ from $\sigma'$ is no later than the replay
state after $P$ from $\sigma$ (ideal model).
If $\lean{checkLatLeaf}\ f\ W\ E\ P\ Q = \mathit{true}$, the same holds in
the block model.
(\lean{checkBlock\_sound}, \lean{checkFP\_sound}, \lean{checkLat\_sound},
\lean{checkLatLeaf\_soundK}) \status{Lean-checked}
\end{theorem}

\lean{checkLat\_sound} uses \cref{thm:c1closure} for hoists and merges,
and monotonicity for a fold, whose live arm runs under fewer gate bits.
\lean{checkLatLeaf} is \lean{checkLat} with the conditions of
\cref{thm:c1block}, and its proof uses that theorem.

The command \texttt{qlin tv-export} writes, for one benchmark, the job
``source to normal form'' and the jobs ``normal form (or its sink) to
fast path'' for $t = 0$ to $6$, the budgets of \cref{sec:eval}. A small
Lean program parses the jobs and
runs the checkers. A normal-form job passes only if \lean{checkBlock} and
\lean{checkLat} both accept it. On all 66 benchmarks, 976 of 976 jobs
pass \status{Lean-checked per output}, so all 66 normal-form outputs are, in the ideal model, no later than its source.
\lean{checkLatLeaf} also accepts all 66, so they are no later in the block model either. The one benchmark whose outcome
records cannot be computed (\texttt{qft-adder-quantum\_7}) contributes
its normal-form job only. A test plants a hoist of an op that one arm lacks, and a
case body that drops the arm its value selects. The checker rejects both.
The checker is sound but not complete: a correct output that it cannot
match would fail, and no output failed.

\subsection{Simulation checks}

Up to 12 qubits, the equivalence check runs both programs on a
state-vector simulator, from $|0\dots0\rangle$ and from seeded random
product states. It compares the output distribution over the bits. It also
compares $\langle v | \rho_\beta | v \rangle$ on all branches $\beta$ for four seeded random vectors $v$, with extra ancilla qubits traced out. A
check of the bits alone cannot see a $Z$/$X$ swap on an unmeasured qubit,
so the witnesses are needed. The check is sound but not complete, and a
planted-error test shows that it rejects a missing correction. It passed
on the normal forms of all benchmarks with at most 12 qubits, on
\nDeferChecked{} defer search leaves, and on 282 fast-path and sink
variants \status{empirical}.

Past 12 qubits, the check runs on a stabilizer tableau (the
Aaronson--Gottesman representation). A one-qubit non-Clifford gate is
allowed on an unentangled qubit, which the tableau recognizes as a $Z$,
$X$, or $Y$ eigenstate. The qubit then holds its 2-vector until
its next measurement. This covers the error source of the noisy
benchmarks, a rotation followed by a measurement. The check compares bit
distributions and, per outcome, the mixture of canonical stabilizer
states, from $|0\dots0\rangle$ and from seeded inputs with up to three
qubits in $X$ or $Y$ eigenstates. The tableau agrees with the dense
simulator, in branches, probabilities, and states, on 28 benchmarks from
$|0\dots0\rangle$, on 114 product-state runs, and on 300 random Clifford
programs. Four planted sign errors in the tableau all fail the random test \status{empirical}.

\paragraph{Defer beyond \cref{thm:defer}.} The copy form and multi-bit
conditions are outside the Lean rule. Beyond the simulation check, a
Wolfram Language script runs teleportation and
\texttt{long\_range\_cnot3} in classical and deferred form, with a
symbolic input state on \texttt{q0}. Both forms give the same cq state, and a planted wrong defer does not \status{Wolfram-checked, two
instances}. The Pauli table has three independent checks: Lean, Wolfram,
and 96 cases against the simulator.

\section{Evaluation}
\label{sec:eval}

The evaluation answers four questions.
\begin{description}
\item[RQ1] Do the compiled programs, and those of the baselines, agree
  with the source on real benchmarks?
\item[RQ2] When does the fast path lower the modeled expected latency?
\item[RQ3] Is the defer choice stable over the sourced range of device
  parameters?
\item[RQ4] Does the cost model agree with an emulator, with the logic of a
  real controller, and with a device?
\end{description}
\Cref{sec:defer} already compares the defer search with the exhaustive optimum.
\emph{Modeled} numbers come from the timing model of \cref{sec:m0},
\emph{emulated} from the Selene emulator, \emph{RTL-simulated} from the
logic of the QubiC controller without qubits, and \emph{measured} from
ibm\_kingston.

\paragraph{Benchmarks.} We use \nBaseBenchmarks{} programs from three
sources:
\begin{itemize}
\item hand-written: teleportation, a repetition code round, a chain of
  adaptive measurements in measurement-based computation,
\item the jeff structured benchmarks: teleportation,
  repeat-until-success, the inverse quantum Fourier transform (iQFT),
  iterative quantum phase estimation (iQPE), and others,
\item the dynamarq suite~\cite{shirgure2026dynamarq}: repetition,
  five-qubit, and Steane codes with lookup decoders, constant-depth
  long-range CNOT and fan-out, GHZ-state preparation, and iterative
  phase estimation.
\end{itemize} Some QEC programs
have a \texttt{\_noisy} variant with an injected error qubit. Before the
first gate on a syndrome ancilla, a data qubit takes an $X$ error with
probability 0.05, once, and the injection ops cost no time. The
latency, search, and checks need simulation without sampling, so programs above 12
qubits (with copy qubits) are reported as skipped.

\paragraph{Cost presets.} HERON\_LIKE holds starting values. A one-qubit
gate takes 32~ns, a two-qubit gate 68~ns, a measurement 800~ns, and a
reset 1000~ns, with $t_\mathrm{ff}=600$~ns. HERON\_KINGSTON takes
gates, measurement (1760~ns mid-circuit), and reset (2312~ns) from the
ibm\_kingston calibration snapshot, and $t_\mathrm{ff}$
from \citet{shirgure2026dynamarq}. All values have a tagged source line,
and a unit test checks the file against the code. The three-qubit gate
time and $t_\mathrm{branch}$ have no source and are marked assumed.

\paragraph{RQ1: semantics on real programs.} \Cref{tab:baselines} runs BQCP~\cite{fulginiti2026bqcp}
(at most 512 amplitudes and 4 branches),
Qiskit at optimization levels 0 and 3, and
TKET~\cite{sivarajah2020tket} on all benchmarks. A semantics guard compares the outputs with the sources by simulation. BQCP reduces the
number of dynamic conditions on \nBaseBqcpFf{} benchmarks. Qiskit and TKET reduce it on no benchmark. Qiskit O3 raises the number of dynamic ops on
\nBaseQiskitOThreeMoreDyn{} benchmarks, because basis translation adds
gates inside the arms. TKET could not convert 13 programs (conditions
on expressions, \texttt{while} loops, and non-basic gates inside
conditions). One BQCP output failed the guard, on
\texttt{hand/mbqc\_chain4}. There BQCP removes an \texttt{if} on a bit
that both arms of an earlier \texttt{if} measure. The output distribution
differs by up to 0.0177, and Qiskit Aer confirms the difference. A
three-qubit program reproduces it: the source gives $P(c_2 = 1) = 0.497$
and the BQCP output gives $0.000$ (Aer, 100000 shots). In the BQCP
implementation, a measurement inside a branch body marks its qubit as
unknown but keeps the old tracked value of its bit. A later \texttt{if}
on that bit then folds as if the bit were still 0. BQCP proves its
analysis sound on paper~\cite{fulginiti2026bqcp}. The mismatch is in the
code, which is the gap that the simulation check covers for qlin too.
All qlin outputs with at most 12 qubits passed the equivalence check of
\cref{sec:soundness}.

\begin{table}[t]
\centering
\small
\begin{tabular}{lrrrr}
\toprule
pipeline & ran & fewer ff & smaller core & more dyn \\
\midrule
bqcp & 46 & 17 & 17 & 0 \\
qiskit\_o0 & 47 & 0 & 0 & 6 \\
qiskit\_o3 & 47 & 0 & 0 & 6 \\
tket & 34 & 0 & 0 & 0 \\
\bottomrule
\end{tabular}

\caption{Baselines on \nBaseBenchmarks{} benchmarks. ``ran'': the
pipeline produced a program that passed the semantics guard. ``fewer
ff'', ``smaller core'': fewer dynamic conditions, or fewer core leaves,
than the source. ``more dyn'': more ops inside dynamic branches. Three
programs are too large for the guard.}
\label{tab:baselines}
\end{table}

\paragraph{RQ2: when the fast path pays.} The block model charges
$t_\mathrm{branch}$ per controller decision, and a \texttt{switch} takes
one decision (jump table), one per bit (decision tree), or one per compare
(chain), as \cref{sec:fastpath} defines. We take $t_\mathrm{branch}$ from
controllers: QubiC~2.0 runs a control-flow
instruction in at most 7 cycles at 500~MHz, \nSwitchQubicTb~ns~\cite{xu2023qubic},
and QICK evaluates a condition and jumps in
\nSwitchQickTb~ns~\cite{stefanazzi2021qick}. We add the
\nIbmSlope~$\mu$s that we measured on ibm\_kingston (RQ4) and
$t_\mathrm{branch}=0$. The sweep covers \nSwitchBenchmarks{} programs,
up to 12 qubits with the dense simulator and 13 to 22 qubits with the
stabilizer backend, which runs Steane, the repetition codes of distance
7, 9, and 11 with a minimum-weight lookup decoder (63, 255, and 1023
corrections), and the rotated surface code of distance 3 with
minimum-weight lookup decoders for both syndrome types (30 corrections).
All 252 fast-path and sink variants past 12 qubits pass the check. A variant gains when it beats the best baseline, the best of the
source, the normal form, the best defer program, BQCP, and Qiskit O3.
\Cref{tab:switch} counts the programs that gain 10\% or more, and
\cref{tab:switchrows} lists the programs that gain at some point
\status{modeled}.

With $t_\mathrm{branch}=0$ the gains come only from the Pauli sink,
which moves corrections past the unencoding Cliffords: 14.3\% on
\texttt{steane\_noisy}, and on the repetition codes prepared in
$|{+}\rangle$ 9.5\% at distance 5 and 12.2, 14.3, and 16.0\% at
distance 7, 9, and 11. From $|0\dots0\rangle$ with no
injected error all syndromes are 0, so BQCP removes all corrections and
wins on the noiseless codes, the surface codes included. At the QICK cost, \nSwitchQickTreeTen{}
programs gain 10\% or more under a decision tree, and the gain grows with
the distance, from 40.7\% at $d=7$ to 91.2\% at $d=11$. The lookup
decoder has one \texttt{if} per syndrome value, and a serial controller pays for all of them, while the switch needs one decision per syndrome bit. The
noisy surface code prepared in $|1\rangle$ gains 13.8\% at the QICK cost
with a decision tree. Prepared in $|{+}\rangle$ it gains 0.6\%, and
there BQCP is the best baseline. The
switch cost changes the count by at most one program at any branch cost.

\paragraph{Switch or speculation.} The gain on these decoders can come
from the switch, which replaces a chain of \texttt{if}s, or from the
budget $t$, which limits the switch to likely outcomes. A full switch,
the fast path with $t$ equal to the number $m$ of group bits, has a case for all outcomes and never takes the exit edge. Without the sink, the
best fast path with $t<m$ beats the full switch by at most
\nFullTabtreeMax\% under a jump table or a decision tree, on all
\nFullPrograms{} programs whose full switch fits the size limit and at all branch costs \status{modeled}. So the gain comes from the switch.
The source here is a lookup decoder with one \texttt{if} per syndrome
value, which is how dynamarq and our generators write it. A decoder
written as a tree of \texttt{if}s on the syndrome bits would take one
decision per bit, as the tree-dispatched switch does. The budget saves
code instead. On \texttt{repetition11\_0\_noisy} at the QICK cost with
a decision tree, the full switch gains \nFullRepFull\% with
\nFullRepFullCases{} cases, $t=2$ gains \nFullRepTwo\% with
\nFullRepTwoCases{} cases, and $t=4$ gains \nFullRepFour\% with
\nFullRepFourCases{} cases. Under a compare chain a case costs a compare, so a smaller switch is also faster. At the IBM cost the best
fast path beats the full switch by \nFullSteaneMargin\% on
\texttt{steane\_noisy} and by \nFullSurfaceMarginLo{} to
\nFullSurfaceMarginHi\% on the two noisy surface codes. These are the
only \nFullChainWins{} programs where this margin exceeds 1\% and the
program gains 10\% or more.

\paragraph{Noise.} For error probabilities from \nNoisePLo{} to
\nNoisePHi{} instead of 0.05, \nNoiseNoneTenLo{} and \nNoiseQickTenLo{}
programs still gain 10\% or more \status{modeled}. On
\texttt{repetition11\_0\_noisy} the best budget gains \nNoiseRepBestLo{}
to \nNoiseRepBestHi\%, while $t=2$ falls from \nNoiseRepTwoLoP{} to
\nNoiseRepTwoHiP\%.

The fast path does not always win. On the five-qubit code the best defer
program beats all fast paths by 17.1 to 18.9\% at 0 to 42~ns, and the
fast path wins only at the IBM cost. At the IBM cost, on \texttt{repetition3\_1\_noisy}, the defer program removes the branches and is 2.7 times
faster than the best fast path. The exit probability decides the best
$t$ (\cref{tab:budget}). A budget over groups with random bits, such as
the logical readout of the five-qubit code, makes $t=0$ exit 95\% of the
time.

\begin{table}[t]
\centering
\small
\begin{tabular}{lrrrr}
\toprule
controller & $t_\mathrm{branch}$ (ns) & table & tree & chain \\
\midrule
zero cost & 0 & 4 (16.0\%) &  &  \\
QubiC & 14 & 8 (80.2\%) & 7 (79.1\%) & 8 (78.0\%) \\
QICK & 42 & 11 (92.0\%) & 10 (91.2\%) & 10 (89.4\%) \\
IBM & 3450 & 13 (99.8\%) & 13 (98.9\%) & 13 (96.9\%) \\
\bottomrule
\end{tabular}

\caption{Programs, of \nSwitchBenchmarks, whose best fast-path or sink
variant beats the best baseline by 10\% or more, with the largest gain in
parentheses (modeled, block model, HERON\_LIKE values,
$t_\mathrm{ff}=600$~ns). Columns: how a \texttt{switch} is dispatched.
With $t_\mathrm{branch}=0$ the dispatch costs nothing. The table column
is a lower bound for a controller without a jump table, such as the
QubiC design of RQ4.}
\label{tab:switch}
\end{table}

\begin{table}[t]
\centering
\scriptsize
\begin{tabular}{lrrrrr}
\toprule
benchmark & qubits & 0~ns & QubiC, tree & QICK, tree & IBM, chain \\
\midrule
\texttt{dynamarq/five\_qubit\_code} & 11 & $-$18.9\% & $-$17.8\% & $-$17.1\% & 29.0\% \\
\texttt{dynamarq/five\_qubit\_code\_noisy} & 12 & $-$18.6\% & $-$18.0\% & $-$17.3\% & 28.7\% \\
\texttt{dynamarq/repetition3\_1\_noisy} & 6 & 3.1\% & 1.6\% & $-$1.5\% & $-$169.8\% \\
\texttt{dynamarq/repetition5\_0\_noisy} & 10 & $-$2.4\% & 2.6\% & 11.4\% & 80.2\% \\
\texttt{dynamarq/repetition5\_1\_noisy} & 10 & 9.5\% & 9.1\% & 12.1\% & 79.7\% \\
\texttt{dynamarq/steane\_noisy} & 15 & 14.3\% & 13.9\% & 12.5\% & 58.7\% \\
\texttt{qec/repetition11\_0\_noisy} & 22 & $-$1.9\% & 78.7\% & 91.2\% & 96.9\% \\
\texttt{qec/repetition11\_1\_noisy} & 22 & 16.0\% & 79.1\% & 90.7\% & 96.9\% \\
\texttt{qec/repetition7\_0\_noisy} & 14 & $-$2.2\% & 18.3\% & 40.7\% & 90.8\% \\
\texttt{qec/repetition7\_1\_noisy} & 14 & 12.2\% & 26.0\% & 41.6\% & 90.6\% \\
\texttt{qec/repetition9\_0\_noisy} & 18 & $-$2.0\% & 49.0\% & 73.4\% & 94.9\% \\
\texttt{qec/repetition9\_1\_noisy} & 18 & 14.3\% & 52.2\% & 72.6\% & 94.9\% \\
\texttt{qec/surface3\_0\_noisy} & 19 & 7.1\% & 5.4\% & 13.8\% & 70.9\% \\
\texttt{qec/surface3\_1\_noisy} & 19 & $-$4.8\% & $-$8.0\% & 0.6\% & 70.5\% \\
\bottomrule
\end{tabular}

\caption{Gain of the best fast-path or sink variant over the best
baseline, for the programs that gain at some point (modeled, as in
\cref{tab:switch}). A negative gain means a baseline is faster; on the
five-qubit code and on \texttt{repetition3\_1\_noisy} it is the best
defer program.}
\label{tab:switchrows}
\end{table}

\begin{table}[t]
\centering
\small
\begin{tabular}{llrrrrrrr}
\toprule
benchmark & branch cost & $t=0$ & 1 & 2 & 3 & 4 & 5 & 6 \\
\midrule
\texttt{dynamarq/five\_qubit\_code} & 0~ns & 1.01 & 1.01 & 1.01 & 1.01 & 1.01 & 1.01 & -- \\
\texttt{dynamarq/five\_qubit\_code} & QICK, tree & 1.03 & 1.01 & 0.98 & 0.96 & 0.96 & 0.96 & -- \\
\texttt{dynamarq/repetition5\_0\_noisy} & 0~ns & 1.00 & 1.00 & 1.00 & 1.00 & 1.00 & -- & -- \\
\texttt{dynamarq/repetition5\_0\_noisy} & QICK, tree & 0.91 & 0.89 & 0.87 & 0.87 & 0.87 & -- & -- \\
\texttt{dynamarq/repetition5\_1\_noisy} & 0~ns & 1.00 & 1.00 & 1.00 & 1.00 & 1.00 & -- & -- \\
\texttt{dynamarq/repetition5\_1\_noisy} & QICK, tree & 0.92 & 0.90 & 0.88 & 0.88 & 0.88 & -- & -- \\
\texttt{dynamarq/repetition3\_1\_noisy} & 0~ns & 1.00 & 1.00 & 1.00 & -- & -- & -- & -- \\
\texttt{dynamarq/repetition3\_1\_noisy} & QICK, tree & 0.99 & 0.99 & 0.99 & -- & -- & -- & -- \\
\bottomrule
\end{tabular}

\caption{Fast-path expected latency over the source latency, for budgets
$t=0$ to 6 (modeled, block model, HERON\_LIKE values,
$t_\mathrm{ff}=600$~ns). Branch cost 0~ns: $t_\mathrm{branch}=0$ with a
jump table. QICK, tree: \nSwitchQickTb~ns per decision with a decision
tree. A dash marks a budget above the one that already covers all
outcomes.}
\label{tab:budget}
\end{table}

\paragraph{RQ3: sensitivity.} We scored the variants over a grid of sourced
ranges: $t_\mathrm{meas}$ from 1760 to 4000~ns, $t_\mathrm{ff}$ from 224
to 700~ns, and $t_\mathrm{branch}$ from 0 to 1000~ns. ``Defer beats the
normal form'' holds at all 12 $(t_\mathrm{meas}, t_\mathrm{ff})$ points on 9
benchmarks and at no point on 17, so it does not flip inside the sourced
ranges. The fast-path gain depends only on $t_\mathrm{branch}$. The
five-qubit gain at $t_\mathrm{branch}=200$~ns does not hold under
HERON\_KINGSTON at any grid point \status{modeled}.

\paragraph{RQ4: Selene.} The Selene SimpleRuntime executes one operation at a
time and charges no feedforward cost, so it cannot test the schedule. It
can test the per-outcome accounting. On \nSeleneVariants{} program
variants of \nSeleneBenchmarks{} benchmarks, \nSeleneExact{} of
\nSeleneShots{} shots take the time that qlin's per-outcome trace predicts, and \nSeleneVariantsExact{} variants match in all shots. The other shots come from Selene's compile-time gate cancellation.
There qlin predicts 64 or 128~ns more, at most 0.17\% of the shot
\status{emulated}. Selene also gives the same output distribution as
qlin's simulator on all benchmarks with at most 12 qubits, within the
stated sampling tolerance.

\paragraph{RQ4: QubiC.} No available device runs the fast path, so we
simulate the logic of an open controller cycle by cycle. We run the first dispatch group of four codes on the register-transfer logic
of the open-source QubiC distributed processor~\cite{xu2023qubic}
(commit \texttt{c22cce8}): one core, no qubits, in Verilator. This design
jumps only to fixed addresses and has no bitwise AND, so a switch compiles to a compare chain or to a binary search
on the syndrome value, and a jump table is not available. MCMit extends QubiC with a
constant-latency multi-control branch~\cite{giortamis2026mcmit}, which we
do not simulate. The syndrome
arrives on the function-processor interface, and we count cycles from its
arrival to the end of the group, for all syndrome values. Over
\nQubicPoints{} programs and values, the count is a linear function of the
executed instructions to within \nQubicResid{} cycles: \nQubicCond{}
cycles per conditional jump, taken or not, and \nQubicJump{} per
unconditional jump \status{RTL-simulated}. QubiC has no not-equal
comparison, so a compiled \texttt{if} takes a compare and a jump, 20~ns at
500~MHz, and the block model at the cited 14~ns per decision underestimates
the source. Weighted by the syndrome probabilities, the group of the
distance-5 repetition code takes \nQubicRepSource{} cycles as source and
\nQubicRepChain{} as a compare chain at $t=2$. On the five-qubit code it
takes \nQubicFiveSource{} cycles against \nQubicFiveTree{} for a tree. The
worst case is slower: a syndrome outside the budget pays the compares and
then the source \texttt{if}s, \nQubicRepChainWorst{} against
\nQubicRepSourceWorst{} cycles. These are controller cycles, not shot
times. The delivery of the syndrome, the same for all variants, is not included.

\paragraph{RQ4: IBM.} We ran on ibm\_kingston with the free Open Plan,
which grants 600~s of device time per 28 days. The IBM toolchain rejects nested conditionals,
measurements inside conditionals, and \texttt{switch}. So the fast path
cannot run there, and a switch runs as a chain of \texttt{if\_else}
blocks.

\begin{figure}[t]
\centering
\begin{minipage}{0.49\linewidth}
\centering
\includegraphics[width=\linewidth]{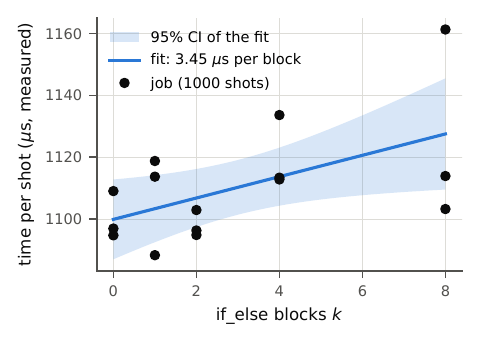}
\end{minipage}\hfill
\begin{minipage}{0.49\linewidth}
\centering
\includegraphics[width=\linewidth]{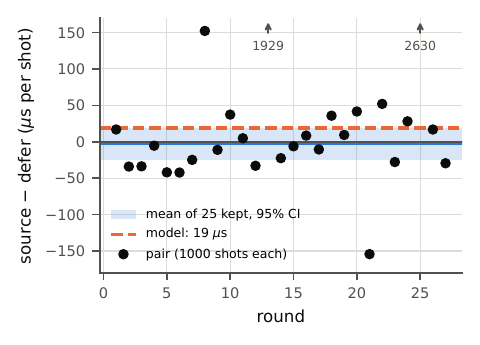}
\end{minipage}
\caption{ibm\_kingston (measured). Left: branch-cost probe, one point per
job, with the least-squares fit and its 95\% band. Right: source minus
best defer program for the five-qubit code, one point per round of the
first run. Two outliers are off the scale (arrows). The band is the mean
and 95\% CI of the 25 pairs kept by a post hoc rule. The 60-round run with
a fixed analysis is not resolved (text). The dashed line is the modeled
gap.}
\label{fig:probe}
\Description{Left: time per shot against the number of if-else blocks, with a fitted line. Right: per-round time difference between source and defer programs, with the modeled gap as a dashed line.}
\label{fig:pairs}
\end{figure}

\emph{Branch cost.} A probe circuit measures one qubit, then runs $k$
\texttt{if\_else} blocks on that bit, then measures a second qubit. We
ran $k \in \{0, 1, 2, 4, 8\}$ with three jobs of 1000 shots per $k$
(\cref{tab:probe}, \cref{fig:probe}). The execution spans give the time
per shot, which includes a fixed repetition delay. The least-squares
slope is \nIbmSlope~$\mu$s per block (95\% CI \nIbmSlopeLo{} to
\nIbmSlopeHi~$\mu$s) \status{measured}. The probe \texttt{if}s read the same bit, so the slope measures $t_\mathrm{branch}$ plus one \texttt{x}
gate per block. It is our only measured value of $t_\mathrm{branch}$.

\emph{Ranking.} At the measured branch cost, the model predicts that the
best defer program of the five-qubit code is 19~$\mu$s per shot faster
than the source. A first run of \nIbmPairs{} rounds, one job per program and round (\cref{fig:pairs}), gave \nIbmKeptMean~$\pm$~\nIbmKeptSe~$\mu$s
(95\% CI \nIbmKeptLo{} to \nIbmKeptHi) after two outlier source jobs were
dropped by a rule chosen after seeing the data. A second run of 60
interleaved rounds, with the analysis and the same outlier rule fixed
before the first job, gave 21.06~$\mu$s (95\% CI $-9.77$ to
51.90~$\mu$s). Against 19~$\mu$s the $t$-test gives $p = 0.89$, and
against 0 the Wilcoxon test gives $p = 0.73$ \status{measured}. The
interval contains both 0 and the predicted 19~$\mu$s, so the run does not
resolve the ranking. The spread between jobs dominates a 19~$\mu$s effect
on a 1.2~ms shot (Shapiro--Wilk $p < 0.0001$ on the differences). The
teleportation variants differ by less than the 7.7~$\mu$s spread of repeated jobs (2 to 7~$\mu$s in the model), so we make no ranking
claim for them.

\begin{figure}[t]
\centering
\includegraphics[width=\linewidth]{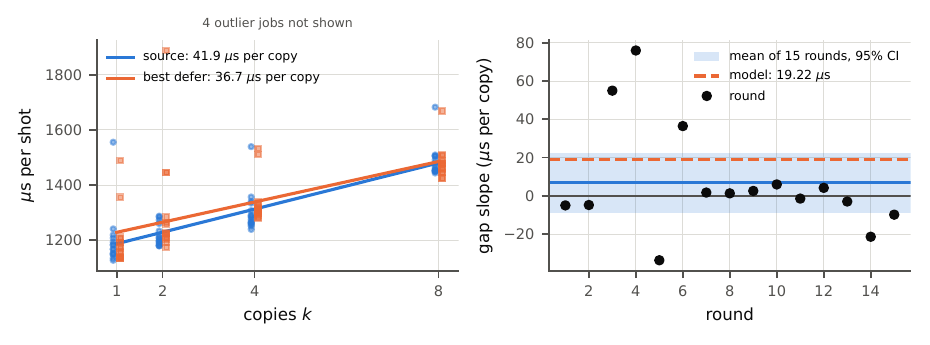}
\caption{Amplified ranking test on ibm\_kingston (measured). Left: time
per shot against the number of copies $k$, one point per kept job, with
a least-squares line per program. The \nCsixbDropped{} jobs that the
outlier rule drops are not shown. Right: the gap slope per round
(source minus best defer program), with its mean and 95\% CI. The
dashed line is the modeled \nCsixbModel~$\mu$s per copy.}
\label{fig:copies}
\Description{Left: time per shot against copies for the source and defer programs, with two nearly parallel fitted lines. Right: per-round gap slopes scattered around a mean near 7 microseconds per copy, with the modeled 19.22 microseconds as a dashed line.}
\end{figure}

\emph{Amplified ranking.} A third run lets the gap grow while the job
noise stays fixed. A program runs its body $k \in \{1, 2, 4, 8\}$
times in one circuit, with all qubits reset between copies, so the
modeled gap grows by \nCsixbModel~$\mu$s per copy. With the analysis
again fixed in advance, \nCsixbRounds{} rounds of one job per program and
$k$ (120 jobs, submitted as one batch, \nCsixbDropped{} dropped by the
outlier rule) give a mean gap slope of \nCsixbMean~$\mu$s per copy (95\%
CI $\nCsixbLo$ to \nCsixbHi~$\mu$s, \cref{fig:copies}). Against
\nCsixbModel~$\mu$s the $t$-test gives $p = \nCsixbTOneNineP$, and against
0 the Wilcoxon test gives $p = \nCsixbWilcoxonP$ \status{measured}. This
run does not resolve the ranking either. Wolfram reproduces the numbers of both analyses from the result files \status{Wolfram-checked}.

\emph{Time per copy (post hoc).} The measured time per copy
(\nCsixbPooledSource{} and \nCsixbPooledDefer~$\mu$s) matches the model
at $t_\mathrm{branch} = \nCsixbEffSource$~$\mu$s for the source and
\nCsixbEffDefer~$\mu$s for the defer program, inside the CI of the probe
and below its point estimate. This reading came after the data.
\Cref{tab:gap} lists the modeled values next to their emulated or measured value.

\begin{table}[t]
\centering
\small
\begin{tabular}{rrrr}
\toprule
if\_else blocks & jobs & mean per shot ($\mu$s) & sd ($\mu$s) \\
\midrule
0 & 3 & 1100.28 & 7.70 \\
1 & 3 & 1106.98 & 16.31 \\
2 & 3 & 1098.07 & 4.30 \\
4 & 3 & 1119.98 & 11.85 \\
8 & 3 & 1126.18 & 30.89 \\
\bottomrule
\end{tabular}

\caption{Branch-cost probe on ibm\_kingston (measured, 1000 shots per
job). The time per shot includes the repetition delay.}
\label{tab:probe}
\end{table}

\begin{table}[t]
\centering
\footnotesize
\begin{tabular}{llllp{3.0cm}}
\toprule
quantity & where & model & measured & gap and mechanism \\
\midrule
time per shot & Selene & per-outcome trace & \nSeleneExact{} of \nSeleneShots{} shots & at most 0.17\% of a shot, from compile-time gate cancellation in Selene \\
$t_\mathrm{branch}$ & IBM & \nSwitchQubicTb, \nSwitchQickTb~ns (FPGA controllers) & \nIbmSlope~$\mu$s (\nIbmSlopeLo{} to \nIbmSlopeHi) & controller time per \texttt{if\_else} block, far above the FPGA controllers \\
decision & QubiC RTL & \nSwitchQubicTb~ns (cited) & 12~ns compare, 20~ns \texttt{if} & no not-equal: an \texttt{if} adds a jump \\
gap per shot & IBM & 19~$\mu$s & 21.06~$\mu$s ($-9.77$ to 51.90) & not resolved: job-to-job spread on a 1.2~ms shot \\
gap per copy & IBM & \nCsixbModel~$\mu$s & \nCsixbMean~$\mu$s ($\nCsixbLo$ to \nCsixbHi) & not resolved \\
time per copy & IBM & \nCsixbModelSource, \nCsixbModelDefer~$\mu$s & \nCsixbPooledSource, \nCsixbPooledDefer~$\mu$s & post hoc: matches a branch cost of \nCsixbEffSource{} and \nCsixbEffDefer~$\mu$s \\
\bottomrule
\end{tabular}
\caption{Modeled against emulated or measured values. Intervals are 95\%
CIs. ``time per copy'': source and best defer program of the five-qubit
code. The model uses the probe's branch cost on IBM.}
\label{tab:gap}
\end{table}

\section{Related Work}
\label{sec:related}

\paragraph{Verified quantum compilation.} VOQC is an optimizer for
quantum circuits written in Coq and extracted to OCaml. Its gate
optimizations are proved correct against the matrix semantics of unitary
SQIR programs, and two optimizations of programs with measurement are
proved against the density-matrix semantics of full
SQIR~\cite{hietala2021voqc}. SQIR has no classical bit store, so a
program branches on an outcome only at its measurement, and VOQC rewrites
gates. qlin keeps the measured bits in its state and rewrites the control
structure. CertiQ and Giallar verify Qiskit
compiler passes with little manual proof, by an SMT encoding and by
symbolic rewrite rules, and both found bugs in the Qiskit
implementation~\cite{shi2019certiq,tao2022giallar}. CoqQ gives a
program logic proved sound against a denotational semantics of a deeply
embedded quantum language~\cite{zhou2023coqq}, and QEDIA mechanizes a
quantum Hoare logic with classical variables in Lean~4 to verify
programs~\cite{ren2026qedia}. Ying et al.\ verify algebraic laws of
quantum programs in Coq and derive the principle of deferred measurement
from them, in a language that discards intermediate
outcomes~\cite{ying2025laws}. The defer rule of qlin keeps the measured
bits in the cq-state (\cref{thm:defer}). A survey covers the
wider field~\cite{lewis2023survey}. QPPM verifies in Lean a compiler
for Pauli-product-measurement programs with classical feedback. Its
compilation and optimizations preserve the outputs and the recorded
measurement results, and its scheduler picks the candidate with the
smallest certified serial-depth bound, checked by a certificate per candidate~\cite{ye2026qppm}. It is the closest verified compiler to
qlin. It proves no theorem that a rewrite
never delays the schedule, and its depth model counts logical
measurements and leaves out Pauli feedback. These works verify that a pass
preserves the semantics. qlin also proves a performance property: hoist
and merge steps never delay the program
(\cref{thm:c1,thm:c1closure,thm:c1block}), and the checker validates this bound for the normal-form outputs. Its
semantics is a cq-state over mid-circuit measurement and classical
control, and its density-matrix model makes the leaf axioms theorems.
VOQC verifies its implementation, so it needs no check per output, and
it trusts Coq extraction and its OCaml driver. qlin uses translation
validation~\cite{pnueli1998tv,necula2000tv} instead: the checker, proved
sound in Lean, validates the semantics of the normal-form and fast-path outputs. Its trusted base moves to the Lean
kernel, the Lean copy of the IR, the exporter in the Rust compiler, and
the job parser. On the programs we compiled, a semantic bug in those
Rust passes would appear as a rejected output, not as a wrong
program. A validator of transpiler
output, also proved sound in Lean, checks a structural policy instead
of equivalence: no gate acts near the qubits of another
tenant~\cite{malik2026confinement}. \Cref{tab:position} compares
what the approaches prove and what they check per output.

\begin{table}[t]
\centering
\footnotesize
\begin{tabular}{@{}*{4}{>{\raggedright\arraybackslash}p{3.15cm}}@{}}
\toprule
approach & semantics & latency & per output \\
\midrule
VOQC~\cite{hietala2021voqc} & optimizer code, Coq & no & not needed (verified code) \\
CertiQ, Giallar~\cite{shi2019certiq,tao2022giallar} & passes, SMT and rewrite rules & no & no \\
QPPM~\cite{ye2026qppm} & compiler, Lean & depth bound & depth certificate \\
BQCP~\cite{fulginiti2026bqcp} & analysis, on paper & no & no \\
qlin & rules, Lean (cq model) & moves, Lean & normal form (with timing), fast path: Lean \\
 & & & sink, defer: simulation \\
\bottomrule
\end{tabular}
\caption{What the approaches prove and check. ``semantics'': what is proved
to preserve the semantics, and how. ``latency'': a proved timing
property of the rewrites (QPPM: a certified depth bound per candidate,
not a theorem about rewrites). ``per output'': a check of the compiled programs. ``no'': the cited work states no such result.}
\label{tab:position}
\end{table}

\paragraph{Classical control in dynamic circuits.} BQCP tracks measured
values and post-measurement
states along bounded paths, and it proves its analysis and its
simplifications sound on paper~\cite{fulginiti2026bqcp}. A related pass
uses symbolic simulation and a probabilistic circuit model to remove
classically controlled operations, and names a hardware-aware cost model
for measurement and feedforward as future
work~\cite{fulginiti2026compile}. Constant propagation across the
classical-quantum boundary can turn quantum operations into classical
ones~\cite{remme2026dequant}, and branch expansion creates
branch-specific simplifications~\cite{chen2025branch}. Chen et al.\
compute measurement outcome probabilities at compile time by constant
propagation, and replace some mid-circuit measurements by randomized
gates~\cite{chen2024probabilistic}. These passes
decide which branches exist. qlin takes the remaining branches as given
and decides how to execute them, against a cost model, with a latency
theorem for its normal form and a mechanized semantics. BQCP has no cost model and does not change how
the remaining branches are dispatched. It can run before qlin, and we
run it as a baseline (\cref{sec:eval}).

\paragraph{Speculation and specialization.} Trace
scheduling~\cite{fisher1981trace} and superblocks~\cite{hwu1993superblock}
compile a likely path as straight-line code with side exits.
Profile-guided code positioning separates hot from cold
code~\cite{pettis1990profile}. Dynamic deoptimization lets optimized code
fall back to unoptimized code by rebuilding its
frames~\cite{holzle1992deopt}. Fl\"uckiger et al.\ give a correctness
argument for speculation guarded by assume instructions with
deoptimization~\cite{fluckiger2018deopt}, and Barri\`ere et al.\ verify
speculation and deoptimization in a JIT compiler in
Coq~\cite{barriere2021deopt}. A verified superblock scheduler in
CompCert proves that scheduling preserves the semantics, and it measures
the speedup~\cite{six2022superblock}. In quantum control, ARTERY predicts
the branch of a feedback in hardware, from branch history and the readout
pulse, and pre-executes instructions~\cite{tian2025artery}. It targets
the feedforward wait, which the fast path does not shorten. The fast path
reduces the number of controller decisions instead, and a predictor could
select a case of its switch early. MCMit adds a constant-latency
multi-control branch instruction to QubiC and lowers the feedback latency
by up to 70\%~\cite{giortamis2026mcmit}. Our cost model has no class for
this instruction. The qlin fast path is a compile-time form
of the split between a likely and an unlikely path, with two
differences. The outcome distribution comes from a noise model, and a
simulation of the source computes it on small programs. The exit
edge sits before any specialized op, so the fallback needs no state
reconstruction. In the terms of verified deoptimization, no speculated
effect precedes the guard and the frame map is empty. This case matters
here because the controller cannot read the quantum state.

\paragraph{Structure and mapping.} Rovara et al.\ compile structured
programs without unrolling them~\cite{rovara2026unrolling}, and our
importer reads such programs in the jeff format. Rice et al.\ make gates
first-class values of an IR~\cite{rice2026dynir}. DynamiQ maps dynamic
circuits onto coupling graphs~\cite{benbetka2026dynamiq}, and CLASS
places qubits to cut the latency between distributed
controllers~\cite{chen2025class}: both change the layout, not the
branches. The Helios runtime maps qubits and schedules transport while
the program runs~\cite{helios2025}, and our cost model has no transport
term. Qubit reuse adds classically controlled operations to save
qubits~\cite{rovara2025reuse,fang2023dynamic}, the opposite of defer.
AC/DC puts feedforward into a synthesis objective~\cite{niu2024acdc}.

\paragraph{Benchmarks and equivalence.} The dynamarq suite defines an
expected depth over outcome probabilities and runs on IBM hardware and on
the Helios-1E emulator~\cite{shirgure2026dynamarq}. We use its programs and its device
numbers. Burgholzer and Wille give two schemes that let equivalence
checkers treat dynamic circuits as if they had no non-unitary
operations~\cite{burgholzer2021nonunitary}. Hybrid path sums give a
symbolic semantics with rewrite rules for programs with classical and
quantum control~\cite{chareton2026hps}. Irene checks the equivalence of
bounded hybrid programs with measurement and classical control, and it
found 15 bugs in quantum compilers~\cite{ke2026irene}. These three could
check the sink outputs and the defer forms outside \cref{thm:defer},
which we check only by simulation (\cref{sec:soundness}). AlchemQ ships an equivalence certificate with its optimized circuits, for unitary circuits~\cite{laabs2026alchemq}.
Neither Irene nor AlchemQ is proved sound in a proof assistant.

\paragraph{Cost of control.} Yuan and Carbin price quantum control flow
in T gates~\cite{yuan2024tcomplexity}. The quantum control machine shows
where classical-style jumps stop working in
superposition~\cite{yuan2024qcm}. Cobble pairs a language with a time and
space cost analysis~\cite{yuan2026cobble}. qlin prices classical control
flow in latency. The defer rewrite converts classical control into
quantum control, so one cost model can choose between them per device.
Palsberg et al.\ ask for IRs with a precise semantics for mid-circuit
measurement, and for space, time, and classical control costs reported
apart~\cite{palsberg2026challenges}. HyQ targets the same latency from
the language side: its staged type system guarantees that all circuits are generated before the program runs, so the controller generates no
circuit between a measurement and the next circuit. Its typing and its
progress proofs are mechanized in Beluga~\cite{sano2026hyq}. This
removes the latency of circuit generation. It does not bound the time
that the controller needs to decide a branch, and HyQ has no cost model
for that time. qlin starts from a program whose circuits exist. It
chooses from the outcome probabilities which outcomes get a specialized
path, and it proves that hoist and merge steps never delay the
schedule.

\paragraph{Common case and fallback in error correction.} Decoders split
a cheap common case from a full fallback: Clique~\cite{ravi2022clique},
hierarchical decoding~\cite{delfosse2020hierarchical}, and speculative
window decoding~\cite{viszlai2025swiper,malarchick2026specqec}. Pipelined fault-tolerant
execution speculates on decoder results at run
time~\cite{spequlation2026}. GreenPeas compiles decoding hypergraphs just
in time, because hypergraphs for all branches cost exponential
memory~\cite{ziad2026greenpeas}. The budget $M_t$ sits between these two
extremes: it precompiles the likely branches and leaves the rest
generic. In a search of arXiv, OpenAlex, Crossref, and Semantic Scholar
(November 2026), we found no earlier
compile-time fast path over measurement outcomes with a sound fallback.

\section{Limitations: Where the Budget Does Not Pay}
\label{sec:limits}

\paragraph{Outcome distributions.} The budget pays only when one
outcome carries most of the probability. Teleportation has two uniform
bits, so its fast path at $t=2$ is the full tree, and its gain comes from
defer. In iQPE and iQFT the outcomes depend on the input phase, and for a phase that the register represents without error BQCP can remove the branches first. From
$|0\dots0\rangle$ with no noise all QEC syndromes are 0, so constant
propagation makes the QEC benchmarks trivial. We read the QEC results as
statements about the \texttt{\_noisy} variants.

\paragraph{Feedforward only.} In the ideal model a branch costs its
feedforward wait and nothing else. The fast path cannot remove that wait:
the switch still waits for its bits. It gains only when the controller
spends time per branch, which is $t_\mathrm{branch}$ in the block model.
With $t_\mathrm{branch}=0$ only the sink gains 10\% or more, on
\nSwitchNoneTableTen{} codes. A real controller can also spend time on
communication between its cores for a branch on bits from several
qubits, which no class of our sweep models.

\paragraph{Scale.} $|M_t|$ grows as $\binom{m}{t}$. On long syndrome
loops $t$ must stay small, or the fast path must work per round. Latency, search, and checks without sampling run up to 12 qubits on any program, and past
12 qubits only on Clifford programs whose non-Clifford gates act on
unentangled qubits. One benchmark (14 qubits) is reported as skipped.

\paragraph{Where the normal form acts.} It changes 3 of the 66 benchmarks
(\texttt{ipe3}, and two that we wrote to exercise it). Its theorems guard
the pass and do not produce the reported gains.

\paragraph{What is not mechanized.} Lean proves the rules, and checkers validate the normal-form and fast-path outputs, but the Rust code itself is not
verified, and the sink and defer outputs are checked by simulation only.
The latency theorems hold in the ideal model and compare two programs on
the same outcome record. In the block model, in which the reported
gains arise, Lean covers moves of single operations only, and the fast
path has no latency theorem in either model.
Two steps are hand-derived: the relaxation bound of the defer search
(checked exhaustively on all prefixes), and the claim that defer can
empty the classical core. \Cref{thm:defer} covers only the
direct form with one control bit and one gate in the arm.

\paragraph{Devices.} The IBM stack cannot run the fast path at all. On
Helios the runtime owns layout and transport~\cite{helios2025}, which our
cost model does not represent. Selene has no feedforward cost. On the
QubiC logic the fast path runs, but without qubits or measurement. On
ibm\_kingston three timing experiments left the ranking of the
five-qubit variants unresolved. So no scheduling ranking of this paper
has a device confirmation.

\section{Conclusion}
\label{sec:conclusion}

qlin compiles the time that a dynamic circuit spends on classical
control. Its normal form moves operations out of branches without
delaying the program, in the ideal timing model for all moves and with a serial controller for moves of single operations. qlin checks the normal-form outputs for semantics and timing, and the fast-path outputs for semantics. These properties, the rewrite rules, the
fast-path step, the Pauli sink, and the direct defer rule are proved in
Lean over a density-matrix model. The fast path compiles
the likely outcomes into one multiway branch and falls back to the
unchanged program for the rest, so no compensation code acts on the
quantum state.

The gains depend on the controller's time per branch decision. Without it, only the Pauli sink gains 10\% or more, on \nSwitchNoneTableTen{}
noisy codes. At the \nSwitchQickTb~ns per decision of QICK,
\nSwitchQickTreeTen{} programs gain 10\% or more, up to
\nSwitchQickTreeMax\%, mostly because one switch replaces a chain of
\texttt{if}s. The budget then saves code more than time. On
ibm\_kingston a branch took \nIbmSlope~$\mu$s (95\% CI \nIbmSlopeLo{} to
\nIbmSlopeHi~$\mu$s), and three timing experiments could not separate the
variants. Open problems are a device run of the fast path, a less noisy
timing measurement, a latency theorem for the fast path, and a
mechanized copy form of defer.

\section*{Use of Generative AI}
We used generative AI tools in this work, and we checked their output as
follows. The Aristotle prover~\cite{achim2025aristotle} produced the
proofs of the rule and latency theorems from statements that we wrote
and froze first (\cref{sec:soundness}). An AI coding assistant (Claude,
by Anthropic) helped write the Rust and Python code, the density-matrix
Lean files, the latency closure theorem, the latency checkers, the
block model, and their proofs, and the text of this paper. The Lean kernel
checks all proofs, and a script checks the axioms of the listed theorems. Tests and the simulation checks cover the code. We checked all numbers in the text against the result files, and all references against a fetched arXiv or Crossref record. The author takes
responsibility for all content.

\section*{Data-Availability Statement}
The artifact holds the Rust implementation, the Lean development, the
benchmark programs with their source records, the scripts that produce the tables, figures, and numbers of this paper, and the raw device results
(\url{https://github.com/rylanmalarchick/qlin}). The full certifier suite takes about 8 minutes on
one desktop machine.

\bibliographystyle{unsrtnat}
\bibliography{refs}

\clearpage
\appendix
\section{Claims ledger}
\label{app:ledger}

\begin{center}
\footnotesize
\begin{tabular}{lp{0.52\linewidth}p{0.3\linewidth}}
\toprule
claim & statement & status \\
\midrule
C1 & Hoist and merge steps, at any position and in any sequence, never delay the replay state & Lean-checked, in expectation too, and checked per M0 output by \lean{checkLat} (ideal model, no truncated loop). Block model: Lean-checked for leaf moves, checked per output by \lean{checkLatLeaf}; a hoisted \texttt{if} can add latency (Lean-checked). Old hand argument wrong, rules fixed \\
C2 & With defer, the classical core can shrink to zero & hand-derived, shown on teleportation \\
C3 & The normal-form rules preserve the semantics & Lean-checked (rules, any model of A1, A2; A1, A2 Lean-checked for density matrices). M0 outputs: Lean-checked per output \\
C4 & Branch and bound finds the optimum on small inputs & exhaustive, \nDeferBnbMatch{} of \nDeferCertified{} rows \\
C5 & Expected-latency gain on benchmarks & modeled (HERON\_LIKE; $t_\mathrm{branch}$ from published controllers; three switch costs) \\
C6 & The cost model ranks variants correctly on a device & measured, not resolved (27 pairs post hoc; 60 pairs and an amplified design with $k \le 8$ copies, analysis fixed first) \\
C7 & Defer preserves distribution and state & direct form, one bit: Lean-checked. Other forms: empirical, Wolfram-checked on two instances \\
C8 & Fast-path size stays within $S=4096$ on QEC & empirical: without the sink for $t \le 6$; sink variants of the larger codes reach $S$ at $t \ge 3$ \\
C9 & The relaxed bound is below all completions & hand-derived, exhaustive on all prefixes \\
C10 & The fast path and the sink preserve the semantics & Lean-checked (dispatch step, whole construction, and sink, density-matrix model); fast-path outputs Lean-checked per output (910 jobs); sink outputs: empirical \\
C11 & The Pauli conjugation table & Lean-checked, Wolfram-checked, 96 simulator cases \\
C12 & Per-outcome accounting matches Selene & emulated, \nSeleneExact{} of \nSeleneShots{} shots match \\
C13 & ``Defer beats the normal form'' holds over the sourced ranges & modeled \\
C15 & Fast-path gain past 12 qubits at $t_\mathrm{branch}=0$ & modeled, 12.2--16.0\% on 4 noisy codes; stabilizer check on 252 variants \\
C16 & Dispatch cycles on the QubiC controller logic & RTL-simulated (one core, no qubits); time linear in executed instructions within 1 cycle \\
C14 & IBM per-branch cost & measured, \nIbmSlope~$\mu$s (95\% CI \nIbmSlopeLo--\nIbmSlopeHi) \\
\bottomrule
\end{tabular}
\end{center}

\end{document}